\pdfoutput=1 %arXiv admin added this line
\documentclass[11pt,final,onecolumn]{IEEEtran}

\PassOptionsToPackage{}{tocbibind}
\PassOptionsToPackage{style=ieee}{biblatex}
\usepackage[colorlinks,tables=false,bibtex=true]{preamble}
\graphicspath{{/}}

\makeatletter
	\ifbool{@bibtex}{}{
		\addbibresource{IEEEabrv,UWA,PaperList,ownpub}
	}
\makeatother

\newcommand{\bref}[1]{(\hyperref[#1]{B\ref{#1}})}

\title{Fundamental Limits of Blind Deconvolution\\\partname~I: Ambiguity Kernel}
\author{Sunav~Choudhary,~\IEEEmembership{Student~Member,~IEEE,} and~Urbashi~Mitra,~\IEEEmembership{Fellow,~IEEE}%
		\thanks{This work has been funded in part by the following grants and organizations:~ONR~N00014-09-1-0700, AFOSR~FA9550-12-1-0215, NSF~CNS-0832186, NSF~CNS-1213128 and NSF~CCF-1117896.
		Parts of this paper were presented at the IEEE International Symposium on Information Theory (ISIT), Honolulu, Hawai'i, June 29 - July 4, 2014~\cite{choudhary2014sbdidentifiability} and at the IEEE Global Conference on Signal and Information Processing (GlobalSIP), Atlanta, Georgia, Dec. 3-5, 2014~\cite{choudhary2014subspaceblind}.}%
		\thanks{S.~Choudhary and U.~Mitra are with the Ming Hsieh Department of Electrical Engineering, Viterbi School of Engineering, University of Southern California, Los Angeles CA 90089, USA (email: \href{mailto:sunavcho@usc.edu}{\protect\nolinkurl{sunavcho@usc.edu}}, \href{mailto:ubli@usc.edu}{\protect\nolinkurl{ubli@usc.edu}}).}}

\begin{document}
	\maketitle

	\begin{abstract}
		Blind deconvolution is an ubiquitous non-linear inverse problem in applications like wireless communications and image processing.
		This problem is generally ill-posed, and there have been efforts to use sparse models for regularizing blind deconvolution to promote signal identifiability.
		\partname~I of this two-part paper characterizes the ambiguity space of blind deconvolution and shows unidentifiability of this inverse problem for \emph{almost every} pair of unconstrained input signals.
		The approach involves lifting the deconvolution problem to a rank one matrix recovery problem and analyzing the rank two null space of the resultant linear operator.
		A measure theoretically tight (parametric and recursive) representation of the key rank two null space is stated and proved.
		This representation is a novel foundational result for signal and code design strategies promoting identifiability under convolutive observation models.
		\partname~II of this paper analyzes the identifiability of \emph{sparsity} constrained blind deconvolution and establishes surprisingly strong negative results on scaling laws for the sparsity-ambiguity trade-off.
	\end{abstract}

	\IEEEpeerreviewmaketitle

	\section{Introduction}
		\label{sec:intro}
		\IEEEPARstart{B}{lind} deconvolution is a challenging inverse problem that is ubiquitous in applications of signal processing, control theory and wireless communications like blind image restoration~\cite{kundur1996blind,levin2011understanding}, blind system identification~\cite{meraim1997blind,bai1999blindsystem} and, blind channel estimation and equalization~\cite{johnson1998blind,diamantaras2002blind,vanderveen1998improved,hopgood2003,liu1996recentblind}.
		In the absence of additional constraints, blind deconvolution is known to be ill-posed and each application mentioned above imposes some form of prior knowledge on the underlying signal structures to render the inverse problem better behaved.
		\emph{Sparsity} based models have been used extensively in the past decade to capture hidden signal structures in many applications of interest.
		Prominent examples of the exploitation of sparsity include natural images admitting sparse wavelet domain representations~\cite{donoho2001sparse,lustig2007sparseMRI}, ultra wide band communication channels exhibiting sparsity in Doppler-delay domain representations~\cite{li2007estimation,berger2010sparse}, and user preferences and topic models displaying low-rank structures~\cite{zhou2008large,papadimitriou1998latent} (sparsity in eigenvalue domain).
		While there have been a few attempts at exploiting sparsity priors for blind deconvolution and related problems~\cite{ahmed2012blind,kammoun2010robustness,herrity2008blind,barchiesi2011dictionary,hegde2011sampling,grady2005survey}, there does not appear to be a general characterization of \emph{identifiability} of such sparse models, except with very restrictive constraints.
		For example, \cite{ahmed2012blind} assumes a \emph{single random subspace} signal model as opposed to the more commonly used \emph{union of subspaces} signal model in compressed sensing~\cite{donoho2006compressed}.
		In this two-part paper, we prove some surprising negative results for signal identifiability in the non-sparse and sparsity constrained blind deconvolution problems, even for the single subspace model.
		For a principled approach to studying these negative results, the first part of the paper develops a measure theoretically tight characterization of the ambiguity space of blind deconvolution.
		Because of the simplicity of the linear convolution map, its (non-linear) kernel can be partially parametrized to yield an analytically useful representation that is instrumental in the derivation of our results.
		To the best of our knowledge, this is a novel parameterization that has not appeared elsewhere in literature.
		The second part~\cite{choudhary2014limitsBDsparsity} of this paper shows that the unfavorable sparsity-ambiguity trade-off results on identifiability are a consequence of the kernel of the linear convolution map being of large dimension (of the order of the signal dimension).

		\subsection{Numerical Examples and Intuition}
			\label{sec:examples}
			We present some constructive numerical examples illustrating the nature and source of unidentifiability for blind linear deconvolution.
			The discrete-time blind linear deconvolution problem can be stated as the task of recovering (up to scalar multiplicative ambiguities) the unknown signal pair $\bb{\vec{x}_{\ast}, \vec{y}_{\ast}} \in \setR^{m} \times \setR^{n}$ from the observation of their noise free linear convolution $\vec{z}_{\ast} = \vec{x}_{\ast} \star \vec{y}_{\ast} \in \setR^{m+n-1}$.
			It is intuitive to see that if there exists another pair $\bb{\vec{x}, \vec{y}} \in \setR^{m} \times \setR^{n}$ such that $\vec{x} \star \vec{y} = \vec{x}_{\ast} \star \vec{y}_{\ast}$ but $\vec{x}$ and $\vec{x}_{\ast}$ are not collinear then $\bb{\vec{x}_{\ast}, \vec{y}_{\ast}}$ is \emph{unidentifiable} (information theoretically impossible to recover uniquely).
			Below, we illustrate what can go wrong with blind deconvolution for $\mathcal{K} = \setR^{11} \times \setR^{7}$ (beside scalar multiplicative ambiguities).

			Consider the vectors $\vec{x}_{1}, \vec{x}_{2} \in \setR^{11}$ and $\vec{y}_{1}, \vec{y}_{2} \in \setR^{7}$ assigned as
			\begin{subequations}
				\begin{alignat}{2}
					\vec{x}_{1}	& = \tpose{\bb{1,0,1,0,0,0,0,0,1,0,1}},	& \quad \vec{y}_{1}	& = \tpose{\bb{1,0,0,0,1,0,0}}, \\
					\vec{x}_{2}	& = \tpose{\bb{1,0,0,0,0,0,0,0,1,0,0}},	& \quad \vec{y}_{2}	& = \tpose{\bb{1,0,1,0,1,0,1}}.
				\end{alignat}
			\end{subequations}
			Clearly, $\vec{x}_{1}$ and $\vec{x}_{2}$ are linearly independent and a simple calculation reveals that
			\begin{equation}
				\vec{x}_{1} \star \vec{y}_{1} = \vec{x}_{2} \star \vec{y}_{2} = \tpose{\bb{1,0,1,0,1,0,1,0,1,0,1,0,1,0,1,0,0}}
				\label{eqn:common convolved output}
			\end{equation}
			implying unidentifiability of both $\bb{\vec{x}_{1}, \vec{y}_{1}}$ and $\bb{\vec{x}_{2}, \vec{y}_{2}}$ within $\mathcal{K} = \setR^{11} \times \setR^{7}$.
			A more non-trivial numerical example can be constructed using the vectors $\vec{x}_{3}$, $\vec{x}_{4}$, $\vec{y}_{3}$ and $\vec{y}_{4}$ (which are functions of $\vec{x}_{1}$, $\vec{x}_{2}$, $\vec{y}_{1}$ and $\vec{y}_{2}$ above), defined as
			\makeatletter
				\if@twocolumn
					\begin{subequations}
						\label{eqn:numerical example cos sin}
						\begin{align}
							\begin{split}
								\vec{x}_{3}	& = \vec{x}_{1} \cos \frac{\pi}{3} - \vec{x}_{2} \sin \frac{\pi}{3}	\\
												& = \tpose{\bb{-0.366, 0, 0.5, 0, 0, 0, 0, 0, -0.366, 0, 0.5}},
							\end{split}	\\
							\begin{split}
								\vec{y}_{3}	& = \vec{y}_{1} \sin \frac{\pi}{6} - \vec{y}_{2} \cos \frac{\pi}{6}	\\
												& = \tpose{\bb{-0.366, 0, -0.866, 0, -0.366, 0, -0.866}},
							\end{split}	\\
							\begin{split}
								\vec{x}_{4}	& = \vec{x}_{1} \cos \frac{\pi}{6} - \vec{x}_{2} \sin \frac{\pi}{6}	\\
												& = \tpose{\bb{0.366, 0, 0.866, 0, 0, 0, 0, 0, 0.366, 0, 0.866}},
							\end{split}	\\
							\begin{split}
								\vec{y}_{4}	& = \vec{y}_{1} \sin \frac{\pi}{3} - \vec{y}_{2} \cos \frac{\pi}{3}	\\
												& = \tpose{\bb{0.366, 0, -0.5, 0, 0.366, 0, -0.5}}.
							\end{split}
						\end{align}
					\end{subequations}
				\else
					\begin{subequations}
						\label{eqn:numerical example cos sin}
						\begin{align}
							\vec{x}_{3}	& = \vec{x}_{1} \cos \frac{\pi}{3} - \vec{x}_{2} \sin \frac{\pi}{3}
							= \tpose{\bb{-0.366, 0, 0.5, 0, 0, 0, 0, 0, -0.366, 0, 0.5}},	\\
							\vec{y}_{3}	& = \vec{y}_{1} \sin \frac{\pi}{6} - \vec{y}_{2} \cos \frac{\pi}{6}
							= \tpose{\bb{-0.366, 0, -0.866, 0, -0.366, 0, -0.866}},	\\
							\vec{x}_{4}	& = \vec{x}_{1} \cos \frac{\pi}{6} - \vec{x}_{2} \sin \frac{\pi}{6}
							= \tpose{\bb{0.366, 0, 0.866, 0, 0, 0, 0, 0, 0.366, 0, 0.866}},	\\
							\vec{y}_{4}	& = \vec{y}_{1} \sin \frac{\pi}{3} - \vec{y}_{2} \cos \frac{\pi}{3}
							= \tpose{\bb{0.366, 0, -0.5, 0, 0.366, 0, -0.5}}.
						\end{align}
					\end{subequations}
				\fi
			\makeatother
			Clearly, $\vec{x}_{3}$ and $\vec{x}_{4}$ are linearly independent and we observe that
			\makeatletter
				\if@twocolumn
					\begin{equation}
						\begin{split}
							\vec{x}_{3} \star \vec{y}_{3}
							& = \bd{0.134, 0, 0.134, 0, -0.299, 0, 0.134, 0,}	\\
							& \quad \tpose{\db{-0.299, 0, 0.134, 0, -0.299, 0, 0.134, 0, -0.433}}	\\
							& = \vec{x}_{4} \star \vec{y}_{4}
						\end{split}
					\end{equation}
				\else
					\begin{equation}
						\begin{split}
							\vec{x}_{3} \star \vec{y}_{3}
							& = \tpose{\bb{0.134, 0, 0.134, 0, -0.299, 0, 0.134, 0, -0.299, 0, 0.134, 0, -0.299, 0, 0.134, 0, -0.433}}	\\
							& = \vec{x}_{4} \star \vec{y}_{4}
						\end{split}
					\end{equation}
				\fi
			\makeatother
			implying unidentifiability of both pairs $\bb{\vec{x}_{3}, \vec{y}_{3}}$ and $\bb{\vec{x}_{4}, \vec{y}_{4}}$ within $\mathcal{K} = \setR^{11} \times \setR^{7}$.

			\begin{figure}
				\centering
				\includegraphics[width=0.7\figwidth]{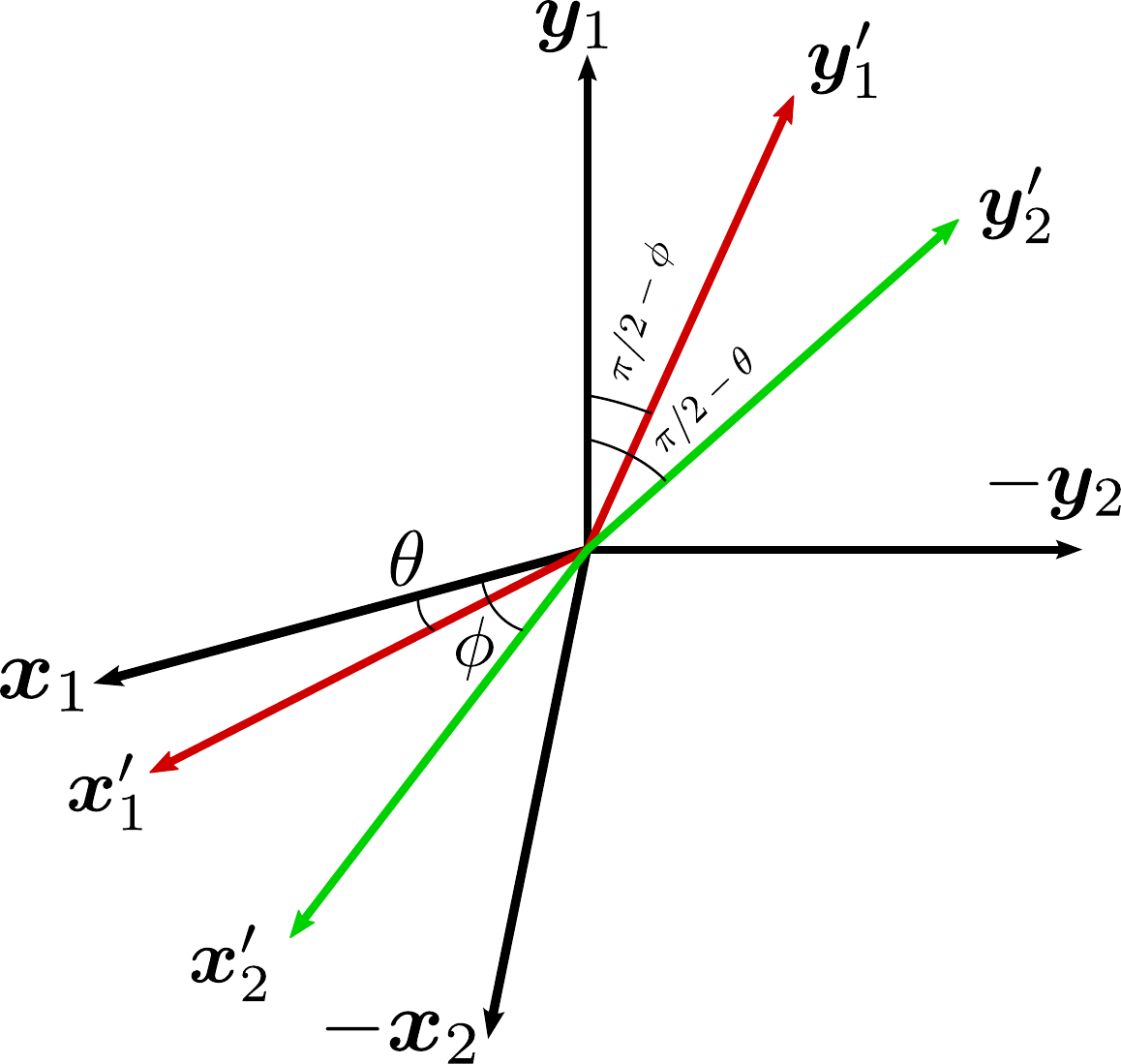}
				\caption{An illustration of rotational ambiguity in blind linear deconvolution.
				Assuming $\vec{x}_{1} \perp \vec{x}_{2}$ and $\vec{y}_{1} \perp \vec{y}_{2}$ in the figure, $\vec{x}_{1} \star \vec{y}_{1} = \vec{x}_{2} \star \vec{y}_{2}$ implies $\vec{x}_{1}' \star \vec{y}_{1}' = \vec{x}_{2}' \star \vec{y}_{2}'$ for any $\theta \neq \phi$.}
				\label{fig:rotation ambiguity}
			\end{figure}
			The construction used in \eqref{eqn:numerical example cos sin} illustrates a form of \emph{rotational ambiguity} that generalizes to produce a family of unidentifiable pairs in $\mathcal{K}$.
			\figurename~\ref{fig:rotation ambiguity} represents a pictorial depiction of the following construction.
			Let $\vec{z}_{0} = \vec{x}_{1} \star \vec{y}_{1} = \vec{x}_{2} \star \vec{y}_{2}$ denote the common convolved output in \eqref{eqn:common convolved output}.
			Consider the parameterized vectors $\vec{x}_{1}', \vec{x}_{2}' \in \setR^{11}$ and $\vec{y}_{1}', \vec{y}_{2}' \in \setR^{7}$ defined by
			\begin{subequations}
				\label{eqn:rotational transform}
				\begin{alignat}{2}
					\vec{x}_{1}'	& = \vec{x}_{1} \cos \theta - \vec{x}_{2} \sin \theta,	& \quad \vec{y}_{1}'	& = \vec{y}_{1} \sin \phi - \vec{y}_{2} \cos \phi,	\\
					\vec{x}_{2}'	& = \vec{x}_{1} \cos \phi - \vec{x}_{2} \sin \phi,	& \quad \vec{y}_{2}'	& = \vec{y}_{1} \sin \theta - \vec{y}_{2} \cos \theta,
				\end{alignat}
			\end{subequations}
			where $\theta \neq \phi$ are the parameters, and $\cc{\vec{x}_{1}, \vec{x}_{2}, \vec{y}_{1}, \vec{y}_{2}}$ acts as the set of seed vectors driving the above transformation.
			Clearly, linear independence of $\vec{x}_{1}$ and $\vec{x}_{2}$ and $\theta \neq \phi$ imply that $\vec{x}_{1}'$ and $\vec{x}_{2}'$ are linearly independent.
			Simple algebraic manipulations reveal that
			\makeatletter
				\if@twocolumn
					\begin{subequations}
						\label{eqn:parametrized example cos sin}
						\begin{align}
							\begin{split}
								\vec{x}_{1}' \star \vec{y}_{1}'
								& =	\vec{x}_{1} \star \vec{y}_{1} \cos \theta \sin \phi + \vec{x}_{2} \star \vec{y}_{2} \sin \theta \cos \phi	\\
								& \quad - \vec{x}_{2} \star \vec{y}_{1} \sin \theta \sin \phi - \vec{x}_{1} \star \vec{y}_{2} \cos \theta \cos \phi	\\
								& = \vec{z}_{0} \sin \bb{\theta + \phi}	\\
								& \quad - \vec{x}_{2} \star \vec{y}_{1} \sin \theta \sin \phi - \vec{x}_{1} \star \vec{y}_{2} \cos \theta \cos \phi
							\end{split}
							\label{eqn:x3 conv y3} \\
							\shortintertext{and,}
							\begin{split}
								\vec{x}_{2}' \star \vec{y}_{2}'
								& =	\vec{x}_{1} \star \vec{y}_{1} \cos \phi \sin \theta + \vec{x}_{2} \star \vec{y}_{2} \sin \phi \cos \theta	\\
								& \quad - \vec{x}_{2} \star \vec{y}_{1} \sin \phi \sin \theta - \vec{x}_{1} \star \vec{y}_{2} \cos \phi \cos \theta	\\
								& = \vec{z}_{0} \sin \bb{\theta + \phi}	\\
								& \quad - \vec{x}_{2} \star \vec{y}_{1} \sin \theta \sin \phi - \vec{x}_{1} \star \vec{y}_{2} \cos \theta \cos \phi.
							\end{split}
							\label{eqn:x4 conv y4}
						\end{align}
					\end{subequations}
				\else
					\begin{subequations}
						\label{eqn:parametrized example cos sin}
						\begin{align}
							\vec{x}_{1}' \star \vec{y}_{1}'
							& =	\vec{x}_{1} \star \vec{y}_{1} \cos \theta \sin \phi + \vec{x}_{2} \star \vec{y}_{2} \sin \theta \cos \phi - \vec{x}_{2} \star \vec{y}_{1} \sin \theta \sin \phi - \vec{x}_{1} \star \vec{y}_{2} \cos \theta \cos \phi	\notag \\
							& = \vec{z}_{0} \sin \bb{\theta + \phi} - \vec{x}_{2} \star \vec{y}_{1} \sin \theta \sin \phi - \vec{x}_{1} \star \vec{y}_{2} \cos \theta \cos \phi	\label{eqn:x3 conv y3} \\
							\shortintertext{and,}
							\vec{x}_{2}' \star \vec{y}_{2}'
							& =	\vec{x}_{1} \star \vec{y}_{1} \cos \phi \sin \theta + \vec{x}_{2} \star \vec{y}_{2} \sin \phi \cos \theta - \vec{x}_{2} \star \vec{y}_{1} \sin \phi \sin \theta - \vec{x}_{1} \star \vec{y}_{2} \cos \phi \cos \theta	\notag \\
							& = \vec{z}_{0} \sin \bb{\theta + \phi} - \vec{x}_{2} \star \vec{y}_{1} \sin \theta \sin \phi - \vec{x}_{1} \star \vec{y}_{2} \cos \theta \cos \phi.	\label{eqn:x4 conv y4}
						\end{align}
					\end{subequations}
				\fi
			\makeatother
			Since \eqref{eqn:x3 conv y3} and \eqref{eqn:x4 conv y4} have the same \rhs, both $\bb{\vec{x}_{1}', \vec{y}_{1}'}$ and $\bb{\vec{x}_{2}', \vec{y}_{2}'}$ are unidentifiable within $\mathcal{K} = \setR^{11} \times \setR^{7}$.
			Since $\bb{\theta, \phi} \in [0, \pi)^{2}$ describes a two dimensional parameter space, \eqref{eqn:parametrized example cos sin} implies that the unidentifiable subset of $\mathcal{K}$ is at least two dimensional.
			Notice that the aforementioned construction for unidentifiable signal pairs (\ie~the rotational ambiguity) remains true, even if the domain of the vectors $\vec{x}_{1}'$ and $\vec{x}_{2}'$ is restricted to an annulus $\vec{x}_{1}', \vec{x}_{2}' \in \set{\vec{w} \in \setR^{11}}{\gamma \leq \twonorm{\vec{w}} \leq \delta}$ for some $\gamma \neq \delta$.

		\subsection{Related Work}
			\label{sec:prior art}
			Prior research on blind system and channel identification~\cite{meraim1997blind,johnson1998blind,liu1996recentblind} has mainly focused on single-in-multiple-out (SIMO) and multiple-in-multiple-out (MIMO) systems, also known as the \emph{blind multi-channel finite-impulse-response (FIR) estimation} problem.
			These systems have multiple output channels that need to be estimated from the observed outputs when the channels are driven by either a single source (SIMO system) or multiple sources (MIMO system).
			A key property necessary for successful identifiability and recovery of the multiple channel vectors is that the channels should display sufficient diversity or richness either stochastically (cyclostationary second order statistics)~\cite{grellier2002analyticalblind,giannakis1998basis,johnson1998blind} or deterministically (no common zero across all channels)~\cite{shin2007blind,xu1995leastsquaresblind,liu1996recentblind,carvalho2004blind,vaidyanathan2004transmultiplexers}.
			As pointed out in~\cite{grellier2002analyticalblind,hopgood2003}, such diversity is generally unavailable in the single-in-single-out (SISO) systems thus making them extremely challenging.
			However, SISO systems are our primary concern due to their equivalence to blind deconvolution problems.
			In this paper, we shall \emph{not} be concerned with stochastic formulations where multiple realizations of the channel are observed to infer empirical statistics.
			Instead, we shall consider non-sparse (here, in \partname~I) and sparse (in \partname~II) channel instances and characterize their instance identifiabilities.
			When blind deconvolution is treated as a channel estimation problem, \cite{manton2003totallyblind} shows that guard intervals of sufficient length between blocks of transmitted symbols enable successful blind channel identification, even for the deterministic SISO system type.
			More specifically, for an $n$ length channel impulse response (CIR) a guard interval of length $n-1$ is needed between consecutive blocks of transmitted source symbols, resulting in an absence of any inter-block or inter-symbol interference (ISI) in the convolved output of the channel.
			We consider scenarios involving ISI and show unidentifiability \wrt~a wide variety of domain restrictions.
			Although our definition of identifiability is similar to that in~\cite{manton2003totallyblind}, our results are substantially different in nature.
			Firstly, we deal with real fields (not algebraically closed) and with sparse subsets of real vector spaces (induced by measures that are not absolutely continuous \wrt~the Lebesgue measure) as opposed to the algebraically closed complex fields and vector spaces considered in \cite{manton2003totallyblind}.
			This requires different proof techniques.
			Secondly, we focus on unidentifiability results and characterize the dimension of the unidentifiable subsets in addition to showing existence results.
			Furthermore, our proofs constructively show unidentifiable subsets as opposed to non-constructive existence proofs.

			To the best of our knowledge, blind deconvolution was cast as a rank one matrix recovery problem first in~\cite{asif2009random}; we adopted this framework in our earlier works~\cite{choudhary2012onidentifiability,choudhary2012identifiabilitybounds,choudhary2013bilinear} on characterization of identifiability in general bilinear inverse problems.
			Herein, we specifically consider the blind deconvolution problem and characterize its inherent \textit{unidentifiability} under some application-motivated sparsity and subspace constraints.
			Further,~\cite{asif2009random} focused on developing heuristic algorithms for the deconvolution problem and did not explicitly address identifiability.
			The subsequent paper~\cite{ahmed2012blind}, by the authors of \cite{asif2009random}, does (implicitly) address identifiability (through a study of recoverability by convex programming) but assumes the knowledge of the support of the sparse signal.
			Although blind deconvolution is an instance of a bilinear inverse problem, the present work differs significantly from our earlier work~\cite{choudhary2013bilinear} in two ways.
			Firstly, \cite{choudhary2013bilinear} focused on identifiability results (analogous to achievability results in information theory) whereas the present paper derives unidentifiability results (akin to converse/impossibility results in information theory).
			Secondly, \partname~II of this paper~\cite{choudhary2014limitsBDsparsity} explores more explicit sparsity priors motivated by applications like cooperative communication~\cite{vajapeyam2008distributed,richard2008sparse}, channel estimation for wide-band communication~\cite{berger2010sparse} and musical source separation~\cite{barchiesi2011dictionary}, whereas \cite{choudhary2013bilinear} developed the theory of regularized bilinear inverse problems more abstractly, making assumptions no more specific than non-convex cone constrained priors.
			A consequence of focusing specifically on blind deconvolution is that we are able to develop tight converse bounds that match our achievability bounds in~\cite{choudhary2013bilinear}.
			The tractability of this analysis hinges on the structural simplicity of the convolution operator after lifting.

			To rigorously justify the dimension related results in both parts of the paper, we call upon the theoretical framework of the \emph{Hausdorff dimension} of sets in metric spaces.
			A moderately detailed introduction to the topic can be found in~\cite{mattila1995geometryofsets}.
			For the sake of readability, we provide a short review of the concept and some of the useful properties in \partname~II, \sectionname~II-C of the paper.

		\subsection{Contributions and Organization}
			\label{sec:contributions}
			In this work, we quantify unidentifiability for certain families of noiseless sparse and non-sparse blind deconvolution problems under a \emph{non-asymptotic} and \emph{non-statistical} setup.
			Specifically, given model orders $m,n \in \setZ^{+}$, \partname~I of the paper develops a parameterization of the ambiguity kernel for blind linear deconvolution and shows that almost every input $\bb{\vec{x}_{\ast}, \vec{y}_{\ast}} \subseteq \setR^{m} \times \setR^{n}$ is unidentifiable under blind deconvolution in the absence of further constraints.
			Furthermore, the characterization developed herein is used in \partname~II of the paper to investigate some application driven choices of the separable domain restriction $\bb{\vec{x}_{\ast}, \vec{y}_{\ast}} \in \mathcal{K} \subseteq \setR^{m} \times \setR^{n}$ and establish that these restrictions are insufficient to guarantee identifiability of the vectors $\vec{x}_{\ast}$ and $\vec{y}_{\ast}$ from their linearly convolved resultant vector $\vec{z}_{\ast} = \vec{x}_{\ast} \star \vec{y}_{\ast}$.
			Our approach leads to the following novelties for \partname~I (refer to~\cite{choudhary2014limitsBDsparsity} for contributions of \partname~II of the paper).

			\begin{enumerate}
				\item	We are able to analyze unidentifiability in blind deconvolution by studying the rank constrained null space of a linear operator on matrices.
						We explicitly demonstrate the almost everywhere unidentifiable nature of unconstrained blind deconvolution by constructing families of adversarial signal pairs for even model orders $m$ and $n$.
						This is a much stronger unidentifiability result than any counterparts in the literature.
				\item	We state and prove a measure theoretically tight, non-linear, partially parametric and partially recursive characterization of the rank two null space of the lifted linear convolution map.
						To the best of our knowledge, this is a new result.
						Characterizing the rank two null space is useful for a variety of applications, including the study of scaling laws in bilinear inverse problems~\cite{choudhary2013bilinear}.
						Our proofs are constructive and demonstrate a rotational ambiguity phenomenon in general bilinear inverse problems which is the cause of existence of a large dimensional set of unidentifiable input pairs in blind deconvolution.
			\end{enumerate}

			The rest of the paper is organized as follows.
			\sectionname~\ref{sec:model} describes the system model, sets up the notion of identifiability up to a suitably defined equivalence class, and presents the lifted reformulation of the blind deconvolution problem as a rank one matrix recovery problem.
			\sectionname~\ref{sec:parametrization} develops a novel, measure theoretically tight characterization of the rank two null space of the lifted linear convolution map, exploiting both parametric as well as recursive definitions.
			\sectionname~\ref{sec:results without coding} presents the key unidentifiability result for non-sparse blind deconvolution, exploiting the parametric representation of the rank two null space of the lifted linear convolution map developed in \sectionname~\ref{sec:parametrization}.
			\sectionname~\ref{sec:conclusion} concludes the paper.
			Detailed proofs of all the results in the paper appear in the \appendicesname~\ref{sec:nullspace converse proof}-\ref{sec:ae unidentifiability proof}.

		\subsection{Notational Conventions}
			\label{sec:notation}
			All vectors are assumed to be column vectors unless stated otherwise.
			We shall use lowercase boldface alphabets to denote column vectors~(\eg~$\vec{a}$) and uppercase boldface alphabets to denote matrices~(\eg~$\mat{A}$).
			The MATLAB\textsuperscript{\circledR} indexing rules will be used to denote parts of a vector/matrix~(\eg~$\mat{A}\bb{2:3,4:6}$ denotes the sub-matrix of $\mat{A}$ formed by the rows $\cc{2,3}$ and columns $\cc{4,5,6}$).
			The all zero vector/matrix shall be denoted by $\vec{0}$ and its dimension would be clear from the usage context.
			For matrices, $\tpose{\bb{\cdot}}$ and $\rank{\cdot}$ respectively return the transpose and rank of their argument.
			Special sets are denoted by uppercase blackboard bold font~(\eg~$\setR$ for real numbers).
			Other sets are denoted by uppercase calligraphic font~(\eg~$\mathcal{S}$).
			Linear operators on matrices are denoted by uppercase script font~(\eg~$\mathscr{S}$).
			For any matrix $\mat{M}$, we denote its column space by $\mathcal{C}\bb{\mat{M}}$.
			The standard Euclidean inner product on a vector space will be denoted by $\ip{\cdot}{\cdot}$ and the underlying vector space will be clear from the usage context.

			When we refer to the `dimension' of a set, we shall mean the Hausdorff dimension of the set.
			To avoid unnecessarily heavy notation, we shall adopt the following convention: The use of both vector variables (like $\vec{x}$, $\vec{y}$, $\vec{u}$, $\vec{v}$, \etc) as well as matrix variables (like $\mat{X}$, $\mat{Y}$, \etc) is allowed to differ across theorems and proofs (and even across different subparts of the same proof) with the understanding that their meanings are restricted to the scope of the individual theorems (and proofs).
			This facilitates the reuse of the same variable names across different theorems with distinct meanings without confusion.

	\section{System Model}
		\label{sec:model}
		\subsection{The Blind Deconvolution Problem}
			\label{sec:convolution model}
			We shall consider the noiseless linear convolution system model
			\begin{equation}
				\vec{z} = \vec{x} \star \vec{y},
				\label{eqn:model}
			\end{equation}
			where $\bb{\vec{x}, \vec{y}}$ denotes the pair of unknown signals with a given domain restriction $\bb{\vec{x}, \vec{y}} \in \mathcal{K} \subseteq \setR^{m} \times \setR^{n}$, $\star \colon \setR^{m} \times \setR^{n} \to \setR^{m + n - 1}$ denotes the linear convolution map, and $\vec{z} \in \setR^{m+n-1}$ is the vector of observations given by
			\begin{equation}
				\vec{z}\bb{l} =	\begin{cases}
									{\displaystyle \sum_{j = 1}^{\min\bb{l,m}} \vec{x}\bb{j} \vec{y}\bb{l+1-j}},	&	1 \leq l \leq n,	\\
									{\displaystyle \sum_{j = l+1-n}^{\min\bb{l,m}} \vec{x}\bb{j} \vec{y}\bb{l+1-j}},	&	1 \leq l-n \leq m-1.
								\end{cases}
				\label{eqn:conv defn}
			\end{equation}
			We are interested in solving for the vectors $\vec{x}$ and $\vec{y}$ from the noiseless observation $\vec{z}$ in~\eqref{eqn:model}.
			The blind linear deconvolution problem corresponding to \eqref{eqn:model} is represented by the feasibility problem
			\find{\bb{\vec{x}, \vec{y}}}
			{\vec{x} \star \vec{y} = \vec{z}, \sep \bb{\vec{x}, \vec{y}} \in \mathcal{K}.}
			{\label{prob:find_xy}}

			We are interested in whether the pair $\bb{\vec{x}, \vec{y}}$ can be uniquely identified in a meaningful sense.
			We assume that the lengths (or model orders) $m$ and $n$, respectively, of vectors $\vec{x}$ and $\vec{y}$ are fixed and known \textit{a priori}.
			Notice that the deconvolution problem~\eqref{prob:find_xy} has an inherent scaling ambiguity due to the identity
			\begin{equation}
				\vec{x} \star \vec{y} = \alpha \vec{x} \star \frac{1}{\alpha} \vec{y}, \quad \forall \alpha \neq 0,
				\label{eqn:scaling ambiguity}
			\end{equation}
			stemming from the bi-linearity of the convolution operator.
			Thus, any meaningful definition of identifiability for blind deconvolution must disregard this type of scaling ambiguity.
			This leads us to the following definition of identifiability.

			\begin{definition}[Identifiability]
				\label{defn:identifiability}
				A vector pair $\bb{\vec{x}, \vec{y}} \in \mathcal{K} \subseteq \setR^{m} \times \setR^{n}$ is identifiable with respect to the linear convolution map~$\star$, if $\forall \bb{\vec{x}', \vec{y}'} \in \mathcal{K} \subseteq \setR^{m} \times \setR^{n}$ satisfying $\vec{x} \star \vec{y} = \vec{x}' \star \vec{y}'$, $\exists \alpha \neq 0$ such that $\bb{\vec{x}', \vec{y}'} = \bb{\alpha \vec{x}, \frac{1}{\alpha} \vec{y}}$.
			\end{definition}

			This definition is in the same spirit as the notion of noise free identifiability described in~\cite{manton2003totallyblind}, but restricted to a set $\mathcal{K} \subseteq \setR^{m} \times \setR^{n}$.
			It is easy to see that \definitionname~\ref{defn:identifiability} induces an equivalence structure on the set of identifiable pairs in $\mathcal{K}$, and identifiability refers to the identification of the equivalence class in $\mathcal{K}$ that generated the observation $\vec{z}$ in \eqref{eqn:model}.

		\subsection{Lifting}
			\label{sec:lifting}
			While \problemname~\eqref{prob:find_xy} is an accurate representation of a blind deconvolution problem, it is not easily amenable to an identifiability analysis in the sense of \definitionname~\ref{defn:identifiability}.
			We use the \emph{lifting} technique from optimization~\cite{balas2005projection} to rewrite \problemname~\eqref{prob:find_xy} as a rank minimization problem subject to linear equality constraints~\cite{choudhary2012onidentifiability,asif2009random}; a form that is better suited for an identifiability analysis,
			\minimize{\mat{W}}
			{\rank{\mat{W}}}
			{\mathscr{S}\bb{\mat{W}} = \vec{z}, \sep \mat{W} \in \mathcal{W},}
			{\label{prob:rank}}
			where $\mathcal{W} \subseteq \setR^{m \times n}$ is \emph{any} set satisfying
			\begin{equation}
				\mathcal{W} \bigcap \set{\mat{W} \in \setR^{m \times n}}{\rank{\mat{W}} \leq 1}
				= \set{\vec{x} \tpose{\vec{y}}}{\bb{\vec{x}, \vec{y}} \in \mathcal{K}},
				\label{eqn:set change}
			\end{equation}
			and $\mathscr{S} \colon \setR^{m \times n} \to \setR^{m+n-1}$ is a linear operator which can be deterministically constructed from the linear convolution map.
			We shall refer to $\mathscr{S}$ as the \emph{lifted linear convolution map}.
			Specifically, $\mathscr{S}\bb{\cdot}$ is the unique linear operator that satisfies
			\begin{equation}
				\mathscr{S}\bb{\vec{x} \tpose{\vec{y}}} = \vec{x} \star \vec{y}, \quad \forall \bb{\vec{x}, \vec{y}} \in \setR^{m} \times \setR^{n}.
				\label{eqn:lifted op}
			\end{equation}
			By construction, the optimal solution to \problemname~\eqref{prob:rank} is a rank one matrix $\mat{W}_{\opt}$ and its singular value decomposition $\mat{W}_{\opt} = \sigma_{\opt} \vec{u}_{\opt} \tpose{\vec{v}_{\opt}}$ yields a solution $\bb{\vec{x}, \vec{y}}_{\opt} = \bb{\sqrt{\sigma_{\opt}} \vec{u}_{\opt}, \sqrt{\sigma_{\opt}} \vec{v}_{\opt}}$ to \problemname~\eqref{prob:find_xy}.
			The exact steps involved in the lifting technique and a proof of equivalence between the lifted and the original problems, in the much broader context of bilinear inverse problems, are available in~\cite{choudhary2013bilinear}.
			Our unidentifiability results in \sectionname~\ref{sec:results without coding} will be based on an analysis of \problemname~\eqref{prob:rank}.

\begin{figure}
				\centering
				\includegraphics[width=0.8\figwidth]{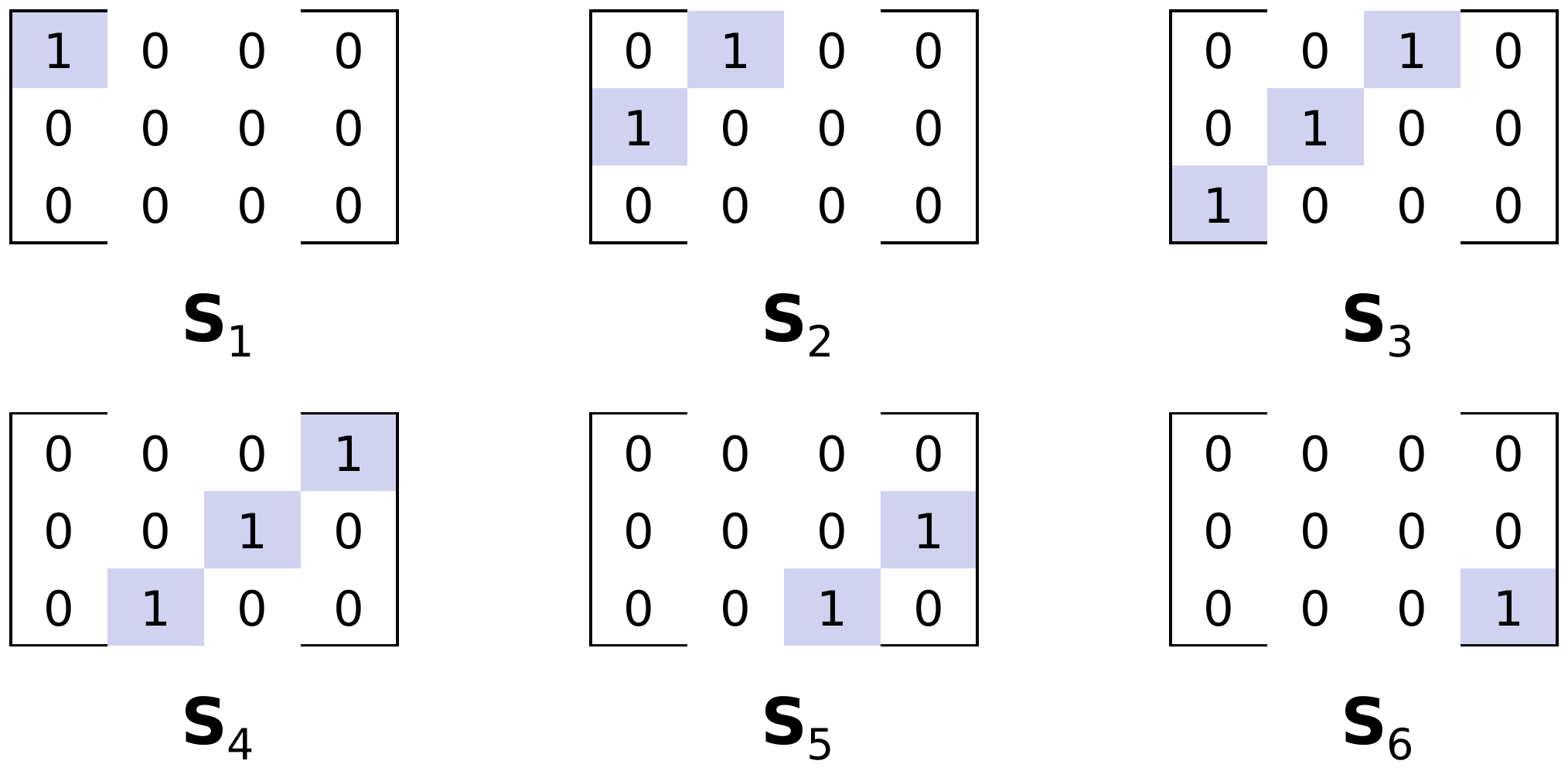}
				\caption{Lifted matrices $\mat{S}_{k} \in \setR^{m \times n}$ for linear convolution map with $m = 3$, $n = 4$ and $1 \leq k \leq m+n-1$.}
				\label{fig:lifting example}
			\end{figure}

			\begin{remark}
				\label{rem:projections}
				It is well known from functional analysis~\cite{helmberg2008introductiontospectral} that any finite dimensional linear operation can be decomposed into a set of inner product operations that collectively define the linear operation.
				The lifted linear convolution map $\mathscr{S}\bb{\cdot}$ can be decomposed into a functionally equivalent set, comprising of $\bb{m+n-1}$ matrices, using coordinate projections.
				Let $\mat{S}_{j} \in \setR^{m \times n}$ denote the $j^{\text{th}}$ matrix in the decomposition and $\phi_{j} \colon \setR^{m+n-1} \to \setR$ denote the $j^{\text{th}}$ coordinate projection operator of $\bb{m+n-1}$ dimensional vectors to scalars, \ie~if $\vec{z} \in \setR^{m+n-1}$ then $\phi_{j}\bb{\vec{z}} = \vec{z}\bb{j}$, for $1 \leq j \leq m+n-1$.
				For all $\bb{\vec{x}, \vec{y}} \in \setR^{m} \times \setR^{n}$, we have the relation
				\makeatletter
					\if@twocolumn
						\begin{equation}
							\begin{split}
								\phi_{j}\bb{\vec{x} \star \vec{y}}
								& = \phi_{j} \circ \mathscr{S}\bb{\vec{x} \tpose{\vec{y}}}
								= \ip{\mat{S}_{j}}{\vec{x} \tpose{\vec{y}}}	\\
								& = \tpose{\vec{x}} \mat{S}_{j} \vec{y}, \quad \forall 1 \leq j \leq m+n-1,
							\end{split}
							\label{eqn:S_j defn}
						\end{equation}
					\else
						\begin{equation}
							\phi_{j}\bb{\vec{x} \star \vec{y}}
							= \phi_{j} \circ \mathscr{S}\bb{\vec{x} \tpose{\vec{y}}}
							= \ip{\mat{S}_{j}}{\vec{x} \tpose{\vec{y}}}
							= \tpose{\vec{x}} \mat{S}_{j} \vec{y}, \quad \forall 1 \leq j \leq m+n-1,
							\label{eqn:S_j defn}
						\end{equation}
					\fi
				\makeatother
				where $\ip{\cdot}{\cdot}$ denotes the trace inner product in the space of matrices $\setR^{m \times n}$.
				By way of exposition, we compute $\mat{S}_{3}$ for $\bb{m,n} = \bb{3,4}$.
				Let $\bb{\vec{x}, \vec{y}} \in \setR^{3} \times \setR^{4}$.
				Invoking the expansion in \eqref{eqn:conv defn}, we have
				\makeatletter
					\if@twocolumn
						\begin{equation}
							\begin{split}
								\phi_{3}\bb{\vec{x} \star \vec{y}}
								& = \sum_{j = 1}^{3} \vec{x}\bb{j} \vec{y}\bb{4-j}	\\
								& =	\BB{\vec{x}\bb{1}, \vec{x}\bb{2}, \vec{x}\bb{3}}
									\begin{bmatrix}
										0	&	0	&	1	&	0	\\
										0	&	1	&	0	&	0	\\
										1	&	0	&	0	&	0
									\end{bmatrix}
									\begin{bmatrix}
										\vec{y}\bb{1}	\\
										\vec{y}\bb{2}	\\
										\vec{y}\bb{3}	\\
										\vec{y}\bb{4}
									\end{bmatrix}	\\
								& =	\tpose{\vec{x}} \mat{S}_{3} \vec{y}.
							\end{split}
						\end{equation}
					\else
						\begin{equation}
							\phi_{3}\bb{\vec{x} \star \vec{y}} = \sum_{j = 1}^{3} \vec{x}\bb{j} \vec{y}\bb{4-j}
							=	\begin{bmatrix}
									\vec{x}\bb{1}	&	\vec{x}\bb{2}	&	\vec{x}\bb{3}
								\end{bmatrix}
								\begin{bmatrix}
									0	&	0	&	1	&	0	\\
									0	&	1	&	0	&	0	\\
									1	&	0	&	0	&	0
								\end{bmatrix}
								\begin{bmatrix}
									\vec{y}\bb{1}	\\
									\vec{y}\bb{2}	\\
									\vec{y}\bb{3}	\\
									\vec{y}\bb{4}
								\end{bmatrix}
							=	\tpose{\vec{x}} \mat{S}_{3} \vec{y}.
						\end{equation}
					\fi
				\makeatother
				The matrix $\mat{S}_{3} \in \setR^{3 \times 4}$ is given by
				\begin{equation}
					\mat{S}_{3}\bb{k,l} =	\begin{cases}
												1,	& k + l = 4, \\
												0,	& \text{otherwise,}
											\end{cases}
					\label{eqn:S_3 expression}
				\end{equation}
				for $1 \leq k \leq 3$ and $1 \leq l \leq 4$.
				In general, it is not hard to see that the matrices $\mat{S}_{j}$, $1 \leq j \leq m+n-1$, are Hankel matrices in $\cc{0,1}^{m \times n} \subset \setR^{m \times n}$ specified as
				\begin{equation}
					\mat{S}_j\bb{k,l} =	\begin{cases}
												1,	& k + l = j + 1, \\
												0,	& \text{otherwise,}
											\end{cases}
					\label{eqn:S_j expression}
				\end{equation}
				for $1 \leq k \leq m$, and $1 \leq l \leq n$.
				\figurename~\ref{fig:lifting example} illustrates these matrices forming the decomposition corresponding to the lifted linear convolution operator $\mathscr{S}\bb{\cdot}$ for the case of $\bb{m, n} = \bb{3, 4}$.
			\end{remark}

		\subsection{Anti-Diagonal Sum Interpretation}
			\label{sec:anti-diagonal sum interpretation}
\begin{figure}
				\centering
				\includegraphics[width=0.6\figwidth]{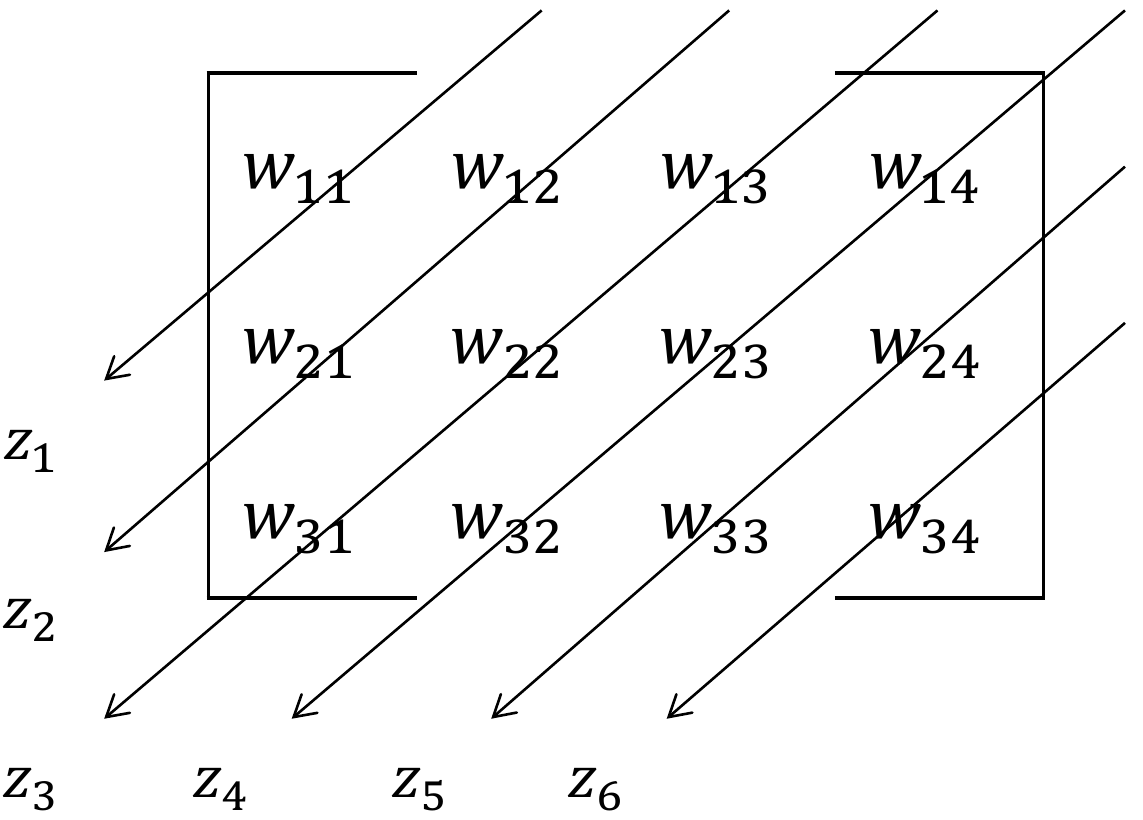}
				\caption{Illustration of the anti-diagonal sum interpretation of the lifted linear convolution operator $\mathscr{S}\bb{\cdot}$ for $\bb{m,n} = \bb{3,4}$ satisfying $\mathscr{S}\bb{\mat{W}} = \vec{z}$.
				The shorthand $w_{kl} = \mat{W}\bb{k,l}$ has been used with $1 \leq k \leq m = 3$ and $1 \leq l \leq n = 4$.}
				\label{fig:anti-diagonal sum}
			\end{figure}

			We illustrate an interpretation of the lifted linear convolution operator that forms a key step in the proofs of our results in \sectionname~\ref{sec:parametrization}.
			Let $\vec{z} \in \setR^{m+n-1}$ be the result of applying the lifted convolution map $\mathscr{S}\bb{\cdot}$ to the matrix $\mat{W} \in \setR^{m \times n}$, \ie~$\mathscr{S}\bb{\mat{W}} = \vec{z}$ and consider the case $\bb{m, n} = \bb{3, 4}$ in \figurename~\ref{fig:anti-diagonal sum} as an example.
			It is clear that $\vec{z}\bb{j}$ is formed by adding all elements of $\mat{W}$ that lie on the $j^{\thp}$ arrow (the $j^{\thp}$ anti-diagonal) in \figurename~\ref{fig:anti-diagonal sum}.
			In other words, \emph{the linear operator $\mathscr{S}\bb{\cdot}$ sums elements of its argument along the anti-diagonals to generate the output}.
			From the definition of linear convolution in \eqref{eqn:conv defn}, it is easy to see that this interpretation of the lifted convolution operator holds regardless of the numerical values of the dimensions $m$ and $n$.

			As an immediate consequence, we make the following observation (also an important part of the proofs to the results in \sectionname~\ref{sec:parametrization}).
			Let $\mat{W}' \in \setR^{\bb{m-1} \times \bb{n-1}}$ be any arbitrary matrix and we consider the matrices $\mat{W}_{1}, \mat{W}_{2} \in \setR^{m \times n}$ constructed as
			\begin{equation}
				\mat{W}_{1} =	\begin{bmatrix}
									\vec{0}	&	\mat{W}'	\\
									0	&	\tpose{\vec{0}}
								\end{bmatrix},
				\quad \mat{W}_{2} =	\begin{bmatrix}
										\tpose{\vec{0}}	&	0	\\
										\mat{W}'	&	\vec{0}
									\end{bmatrix}.
				\label{eqn:block shifted construction}
			\end{equation}
			Clearly, the elements of $\mat{W}_{2}$ along the $j^{\thp}$ anti-diagonal are a circularly shifted copy of the elements of $\mat{W}_{1}$ along the $j^{\thp}$ anti-diagonal, \ie~$\mat{W}_{2}$ is formed by shifting down the elements of $\mat{W}_{1}$ by one unit along the anti-diagonals.
			In particular, this implies that the sum of elements along the $j^{\thp}$ anti-diagonal is the same for both $\mat{W}_{1}$ and $\mat{W}_{2}$.
			Using the \emph{sum along anti-diagonals} interpretation of the lifted linear convolution map $\mathscr{S}\bb{\cdot}$, we have
			\begin{equation}
				\mathscr{S}\bb{\mat{W}_{1}} = \mathscr{S}\bb{\mat{W}_{2}} =	\begin{bmatrix}
																				0	\\
																				\mathscr{S}'\bb{\mat{W}'}	\\
																				0
																			\end{bmatrix},
				\label{eqn:equality under convolution}
			\end{equation}
			where $\mathscr{S}'\fcolon \setR^{(m-1) \times (n-1)} \to \setR^{m+n-3}$ denotes the lifted linear convolution map in one lower dimension than $\mathscr{S}\bb{\cdot}$ \wrt~both rows and columns.
			Thus, $\mat{W}_{1}$ and $\mat{W}_{2}$ are indistinguishable under the action of the linear operator $\mathscr{S}\bb{\cdot}$.

	\section{Parameterizing the Rank Two Null Space}
		\label{sec:parametrization}
		Suppose that $\mathscr{S}\bb{\cdot}$ is the lifted linear operator corresponding to the linear convolution map.
		We denote the rank two null space of $\mathscr{S}\bb{\cdot}$ by $\mathcal{N}\bb{\mathscr{S}, 2}$ which is defined as~\cite{choudhary2013bilinear}
		\begin{equation}
			\mathcal{N}\bb{\mathscr{S}, 2} \triangleq \set{\mat{Q} \in \setR^{m \times n}}{\rank{\mat{Q}} \leq 2,\, \mathscr{S}\bb{\mat{Q}} = \vec{0}}.
		\end{equation}
		In this section, we establish a partially parametric and partially dimension recursive characterization of $\mathcal{N}\bb{\mathscr{S}, 2}$ that is critical to our subsequent results in \sectionname~\ref{sec:results without coding}.
		We first establish \lemmaname~\ref{lem:rank-2 nullspace} that describes a subset of $\mathcal{N}\bb{\mathscr{S}, 2}$.
		\propositionname~\ref{prop:rank-2 nullspace converse} shows existence of subsets of $\mathcal{N}\bb{\mathscr{S}, 2}$ not covered by \lemmaname~\ref{lem:rank-2 nullspace}.
		\theoremname~\ref{thm:rank-2 nullspace full} then provides an almost everywhere characterization of $\mathcal{N}\bb{\mathscr{S}, 2}$.
		Finally, \propositionname~\ref{prop:rank-2 nullspace exception set} shows existence of subsets of $\mathcal{N}\bb{\mathscr{S}, 2}$ that are not covered by \theoremname~\ref{thm:rank-2 nullspace full}.
		We shall make use of the fact that the rank one null space of the linear convolution operator is trivial, \ie
		\begin{equation}
			\mathcal{N}\bb{\mathscr{S}, 1} \triangleq \set{\mat{Q} \in \setR^{m \times n}}{\rank{\mat{Q}} \leq 1,\, \mathscr{S}\bb{\mat{Q}} = \vec{0}} = \cc{\mat{0}},
		\end{equation}
		which follows from interpreting convolution as polynomial multiplication, since the product of two real polynomials is identically zero if and only if at least one of them is identically zero.

		\lemmaname~\ref{lem:rank-2 nullspace} enables the analysis of the rank two null space of $\mathscr{S}\bb{\cdot}$, which is equivalent to the non-linear ambiguity space of blind deconvolution in a transformed domain.
		\begin{lemma}
			\label{lem:rank-2 nullspace}
			Let $m,n \geq 2$ and $\mat{Q} \in \setR^{m \times n}$ admit a factorization of the form
			\begin{equation}
				\mat{Q} =	\begin{bmatrix}
								\vec{u}	&	0	\\
								0	&	-\vec{u}
							\end{bmatrix}
							\begin{bmatrix}
								0	&	\tpose{\vec{v}}	\\
								\tpose{\vec{v}}	&	0
							\end{bmatrix},
				\label{eqn:rank-2 nullspace}
			\end{equation}
			for some $\vec{v} \in \setR^{n-1}$ and $\vec{u} \in \setR^{m-1}$.
			Then $\mat{Q} \in \mathcal{N}\bb{\mathscr{S}, 2}$.
		\end{lemma}
		
		\begin{remark}
			In \lemmaname~\ref{lem:rank-2 nullspace} above, we have used the shorthand representations
			\begin{equation}
				\begin{bmatrix}
					\vec{u}	&	0	\\
					0	&	-\vec{u}
				\end{bmatrix}
				=	\begin{bmatrix}
						\vec{u}(1)	&	0	\\
						\vec{u}(2)	&	-\vec{u}(1)	\\
							\vdots	&	\vdots	\\
					\vec{u}\bb{m-1}	&	-\vec{u}\bb{m-2}	\\
								0	&	-\vec{u}\bb{m-1}
				\end{bmatrix},
			\end{equation}
			and
			\begin{equation}
				\begin{bmatrix}
					0	&	\tpose{\vec{v}}	\\
					\tpose{\vec{v}}	&	0
				\end{bmatrix}
				=	\begin{bmatrix}
								0	&	\vec{v}(1)	&	\vec{v}(2)	&	\cdots	&	\vec{v}\bb{n-1}	\\
						\vec{v}(1)	&	\vec{v}(2)	&	\cdots	&	\vec{v}\bb{n-1}	&	0	\\
					\end{bmatrix}.
			\end{equation}
			For brevity, we shall continue to use analogous shorthand representations for matrices in the rest of the paper, with the understanding that these symbolic shorthands are dimension-wise consistent upon element-wise expansion.
		\end{remark}
		
		\begin{IEEEproof}
			Let $\mat{Q}$ admit a factorization as in \eqref{eqn:rank-2 nullspace}.
			Then,
			\begin{equation}
				\mat{Q} =	\underbrace{\begin{bmatrix}
											\vec{0}	&	\vec{u}\tpose{\vec{v}}	\\
											0	&	\tpose{\vec{0}}
										\end{bmatrix}}_{\mat{Q}_{1}}
						+	\underbrace{\begin{bmatrix}
											\tpose{\vec{0}}		&	0	\\
											- \vec{u} \tpose{\vec{v}}	&	\vec{0}
										\end{bmatrix}}_{\mat{Q}_{2}}
				\label{eqn:anti-diagonal representation}
			\end{equation}
			Clearly, $\mat{Q}_{2}$ in \eqref{eqn:anti-diagonal representation} is obtained by shifting down the elements of $\mat{Q}_{1}$ by one unit along the anti-diagonals and then flipping the sign of each element.
			Since the convolution operator $\mathscr{S}\bb{\cdot}$ sums elements along the anti-diagonals (see \figurename~\ref{fig:anti-diagonal sum} for illustration and \sectionname~\ref{sec:anti-diagonal sum interpretation} for details), the representation of $\mat{Q}$ as in \eqref{eqn:anti-diagonal representation} immediately implies that $\mathscr{S}\bb{\mat{Q}} = \vec{0}$.
			Since \eqref{eqn:rank-2 nullspace} implies that $\rank{\mat{Q}} \leq 2$ so we have $\mat{Q} \in \mathcal{N}\bb{\mathscr{S}, 2}$.
		\end{IEEEproof}
		
		We notice that $\mathscr{S}\bb{\cdot}$ maps $\setR^{m \times n}$ to $\setR^{m+n-1}$.
		An $m \times n$ dimensional rank two matrix has $2(m+n-2)$ degrees of freedom (DoF), so that $\mathcal{N}\bb{\mathscr{S}, 2}$ has at most $(2m+2n-4) - (m+n-1) = (m+n-3)$ DoF.
		Since the representation on the \rhs~of \eqref{eqn:rank-2 nullspace} also has $(m+n-3)$ DoF and $\mathcal{N}\bb{\mathscr{S}, 1} = \cc{\mat{0}}$, our parametrization is tight up to DoF.
		However, the converse of \lemmaname~\ref{lem:rank-2 nullspace} is false in general, as we show in \propositionname~\ref{prop:rank-2 nullspace converse} below.
		This implies that while unidentifiability results for blind deconvolution can be shown by proving existence of matrices satisfying \eqref{eqn:rank-2 nullspace}, proof of identifiability results for deterministic input signals, on the other hand, requires substantially more mathematical effort and careful analysis on the subset of $\mathcal{N}\bb{\mathscr{S}, 2}$ that is not element-wise representable in the form of \eqref{eqn:rank-2 nullspace}.

		\begin{proposition}
			\label{prop:rank-2 nullspace converse}
			For $m,n \geq 3$, there exists a $(m+n-3)$ dimensional set $\mathcal{M}$ such that $\mathcal{M} \subset \mathcal{N}\bb{\mathscr{S}, 2} \subset \setR^{m \times n}$ and for every $\mat{M} \in \mathcal{M}$, $\mat{M}$ is not representable in the form of \eqref{eqn:rank-2 nullspace}.
		\end{proposition}
		
		\begin{IEEEproof}
			\appendixname~\ref{sec:nullspace converse proof}.
		\end{IEEEproof}

		\propositionname~\ref{prop:rank-2 nullspace converse} shows that \lemmaname~\ref{lem:rank-2 nullspace} only provides a partial characterization of $\mathcal{N}\bb{\mathscr{S}, 2}$.
		In particular, the elements of $\mathcal{N}\bb{\mathscr{S}, 2}$ that are not representable by \eqref{eqn:rank-2 nullspace} form an equally large set.

		We are ready to state an almost everywhere description of $\mathcal{N}\bb{\mathscr{S}, 2}$ for the lifted linear convolution map $\mathscr{S}\fcolon \setR^{m \times n} \to \setR^{m+n-1}$.
		Our description will involve a recursion over dimension.
		To avoid any confusion, we shall explicitly augment the dimension to the symbolic representation of the lifted operator.
		Thus, the lifted linear convolution map $\mathscr{S}\fcolon \setR^{m \times n} \to \setR^{m+n-1}$ will be denoted by $\mathscr{S}_{\bb{m,n}}$ and the associated rank two null space will be denoted by $\mathcal{N}\bb{\mathscr{S}_{\bb{m,n}}, 2}$.
		For $m,n \geq 2$, we define the following sets
		\makeatletter
			\if@twocolumn
				\begin{equation}
					\mathcal{M}(m,n) \triangleq \set{\mat{Q} \in \mathcal{N}\bb{\mathscr{S}_{\bb{m,n}}, 2}}{\mat{Q}\bb{m,1} = \mat{Q}\bb{1,n} = 0},
				\end{equation}
				\begin{equation}
					\mathcal{N}_{0}\bb{m, n} \triangleq \set{	\begin{bmatrix}
																		\vec{u}	&	0	\\
																		0	&	-\vec{u}
																	\end{bmatrix}
																	\begin{bmatrix}
																		0	&	\tpose{\vec{v}}	\\
																		\tpose{\vec{v}}	&	0
																	\end{bmatrix}}
																	{\vec{u} \in \setR^{m-1}, \vec{v} \in \setR^{n-1}},
																	\label{eqn:rank-2 nullspace part 1}
				\end{equation}
				\begin{equation}
					\begin{split}
						\mathcal{N}_{2}\bb{m, n}
						& \triangleq \mleft\{	\begin{bmatrix}
													\vec{u}	&	0	\\
													0	&	\vec{u}_{\ast}
												\end{bmatrix}
												\begin{bmatrix}
													0	&	\tpose{\vec{v}}	\\
													\tpose{\vec{v}_{\ast}}	&	0
												\end{bmatrix} \, \middle| \mright.	\\
						& \qquad \mleft. \vphantom{\begin{bmatrix} \vec{u} & 0 \\ 0 & \vec{u}_{\ast} \end{bmatrix}}
						\vec{u} \tpose{\vec{v}} + \vec{u}_{\ast} \tpose{\vec{v}_{\ast}} \in \mathcal{N}\bb{\mathscr{S}_{\bb{m-1,n-1}}, 2} \setminus \cc{\mat{0}} \mright\}.
					\end{split}
					\label{eqn:rank-2 nullspace part 2}
				\end{equation}
			\else
				\begin{align}
					\mathcal{M}(m,n)	& \triangleq	\set{\mat{Q} \in \mathcal{N}\bb{\mathscr{S}_{\bb{m,n}}, 2}}{\mat{Q}\bb{m,1} = \mat{Q}\bb{1,n} = 0},	\\
					\mathcal{N}_{0}\bb{m, n}	& \triangleq	\set{\begin{bmatrix}
																		\vec{u}	&	0	\\
																		0	&	-\vec{u}
																	\end{bmatrix}
																	\begin{bmatrix}
																		0	&	\tpose{\vec{v}}	\\
																		\tpose{\vec{v}}	&	0
																	\end{bmatrix}}
																	{\vec{u} \in \setR^{m-1}, \vec{v} \in \setR^{n-1}},
																	\label{eqn:rank-2 nullspace part 1}	\\
					\mathcal{N}_{2}\bb{m, n}	& \triangleq	\set{\begin{bmatrix}
																		\vec{u}	&	0	\\
																		0	&	\vec{u}_{\ast}
																	\end{bmatrix}
																	\begin{bmatrix}
																		0	&	\tpose{\vec{v}}	\\
																		\tpose{\vec{v}_{\ast}}	&	0
																	\end{bmatrix}}
																	{\vec{u} \tpose{\vec{v}} + \vec{u}_{\ast} \tpose{\vec{v}_{\ast}} \in \mathcal{N}\bb{\mathscr{S}_{\bb{m-1,n-1}}, 2} \setminus \cc{\mat{0}}}.
																	\label{eqn:rank-2 nullspace part 2}
				\end{align}
			\fi
		\makeatother

		\begin{theorem}
			\label{thm:rank-2 nullspace full}
			The following relationships hold.
			\begin{enumerate}
				\item	\label{itm:m1 n1}
						$\mathcal{N}\bb{\mathscr{S}_{\bb{1,n}}, 2} = \cc{\mat{0}}$ and $\mathcal{N}\bb{\mathscr{S}_{\bb{m,1}}, 2} = \cc{\mat{0}}$ for all positive integers $m,n$.
				\item	\label{itm:m2 n2}
						$\mathcal{N}\bb{\mathscr{S}_{\bb{m,2}}, 2} = \mathcal{N}_{0}\bb{m, 2}$ and $\mathcal{N}\bb{\mathscr{S}_{\bb{2,n}}, 2} = \mathcal{N}_{0}\bb{2,n}$ for integers $m,n \geq 2$.
				\item	\label{itm:m n subset}
						$\mathcal{N}\bb{\mathscr{S}_{\bb{m,n}}, 2} \supseteq \mathcal{N}_{0}\bb{m, n} \bigcup \mathcal{N}_{2}\bb{m, n}$ for integers $m,n \geq 3$.
				\item	\label{itm:m n restricted}
						$\mathcal{N}\bb{\mathscr{S}_{\bb{m,n}}, 2} \setminus \mathcal{M}(m,n) = \bb[\big]{\mathcal{N}_{0}\bb{m, n} \bigcup \mathcal{N}_{2}\bb{m, n}} \setminus \mathcal{M}(m,n)$ for integers $m,n \geq 3$.
			\end{enumerate}
		\end{theorem}

		\begin{IEEEproof}
			\appendixname~\ref{sec:full null space proof}.
		\end{IEEEproof}

		\theoremname~\ref{thm:rank-2 nullspace full} provides an \emph{almost} full characterization of the ambiguity set for blind deconvolution, \ie~beside the constructions used in \lemmaname~\ref{lem:rank-2 nullspace} and \propositionname~\ref{prop:rank-2 nullspace converse}, this ambiguity space contains (measure theoretically) very few elements.
		In fact, for $m=2$ or $n=2$, the characterization provided by \lemmaname~\ref{lem:rank-2 nullspace} is complete.
		Another interpretation of \theoremname~\ref{thm:rank-2 nullspace full} is that it provides a converse characterization of $\mathcal{N}\bb{\mathscr{S}, 2}$ while \lemmaname~\ref{lem:rank-2 nullspace} and \propositionname~\ref{prop:rank-2 nullspace converse} provide achievability/existence characterizations for $\mathcal{N}\bb{\mathscr{S}, 2}$.

		\begin{figure}
			\centering
			\includegraphics[width=0.8\figwidth]{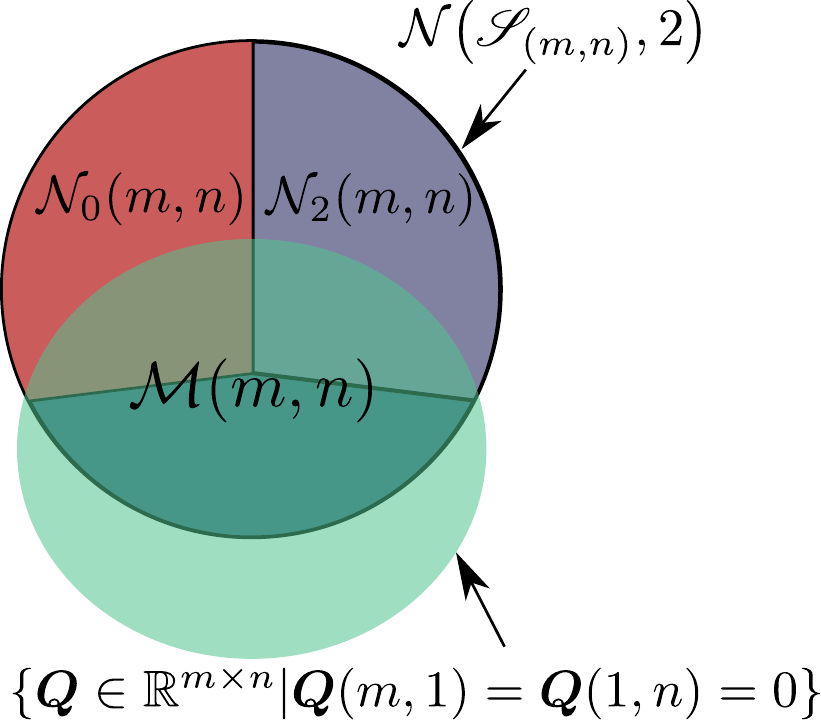}
			\caption{Venn diagram displaying the subset/superset relationships between $\mathcal{N}\bb{\mathscr{S}_{\bb{m,n}}, 2}$, $\mathcal{N}_{0}\bb{m, n}$, $\mathcal{N}_{2}\bb{m, n}$ and $\mathcal{M}\bb{m,n}$ as indicated by \theoremname~\ref{thm:rank-2 nullspace full} for $m,n \geq 3$.}
			\label{fig:nullspace-venn-diagram}
		\end{figure}

		The sets $\mathcal{N}\bb{\mathscr{S}_{\bb{2,n}}, 2} = \mathcal{N}_{0}\bb{2,n}$ and $\mathcal{N}\bb{\mathscr{S}_{\bb{m,2}}, 2} = \mathcal{N}_{0}\bb{m,2}$ are explicitly parametrized $\forall m,n \geq 2$.
		In principle, to construct elements from $\mathcal{N}_{2}\bb{m,n}$ for arbitrary $m, n \geq 3$ with $n \geq m$, we can start from the explicitly parameterized set $\mathcal{N}\bb{\mathscr{S}_{\bb{2,n-m+2}}, 2} = \mathcal{N}_{0}\bb{2,n-m+2}$ and apply \eqref{eqn:rank-2 nullspace part 2} recursively $\bb{m-2}$ times.
		\figurename~\ref{fig:nullspace-venn-diagram} shows a Venn diagram displaying relationships among the different sets in \theoremname~\ref{thm:rank-2 nullspace full}.
		We note that \theoremname~\ref{thm:rank-2 nullspace full} proves equality of the sets $\mathcal{N}\bb{\mathscr{S}_{\bb{m,n}}, 2}$ and $\mathcal{N}_{0}\bb{m, n} \bigcup \mathcal{N}_{2}\bb{m, n}$ when restricted to $\mathcal{M}\bb{m,n}^{\comp}$.
		In \propositionname~\ref{prop:rank-2 nullspace exception set} below, we show that this restriction is necessary when $m=n \geq 3$.
		For $m,n \geq 3$ and $m \neq n$, it is presently unclear whether $\mathcal{M}\bb{m,n}$ is a non-empty set.
		We leave this as an open question.
		Thus, \propositionname~\ref{prop:rank-2 nullspace exception set} characterizes the gap left by the converse characterization in \theoremname~\ref{thm:rank-2 nullspace full} for $m=n \geq 3$.

		\begin{proposition}
			\label{prop:rank-2 nullspace exception set}
			For $n \geq 3$, $\mathcal{N}\bb{\mathscr{S}_{\bb{n,n}}, 2} \bigcap \mathcal{M}\bb{n,n} \setminus \bb[\big]{\mathcal{N}_{0}\bb{n, n} \bigcup \mathcal{N}_{2}\bb{n, n}}$ is a non-empty set of dimension at least $(n-1)$.
		\end{proposition}

		\begin{IEEEproof}
			\appendixname~\ref{sec:nullspace exception set proof}.
		\end{IEEEproof}

		While we shall only use \lemmaname~\ref{lem:rank-2 nullspace} in the sequel, we briefly emphasize the importance of the other results in this section.
		\propositionname~\ref{prop:rank-2 nullspace converse} and \theoremname~\ref{thm:rank-2 nullspace full} supply the intuition that the rank two null space of the linear convolution map is a substantially complex geometrical object.
		Roughly speaking, the rank two null space $\mathcal{N}\bb{\mathscr{S}_{\bb{m,n}}, 2}$ can be divided into exactly two measure theoretically significant (dimension-wise significant) disjoint parts, \viz~$\mathcal{N}_{0}\bb{m, n}$ and $\mathcal{N}_{2}\bb{m, n}$, and a measure theoretically insignificant (not necessarily disjoint from $\mathcal{N}_{0}\bb{m, n} \bigcup \mathcal{N}_{2}\bb{m, n}$) part $\mathcal{M}\bb{m,n}$.
		Both these parts are equally significant since they are of the same dimension, and whatever part of $\mathcal{N}\bb{\mathscr{S}_{\bb{m,n}}, 2}$ is outside $\mathcal{N}_{0}\bb{m, n} \bigcup \mathcal{N}_{2}\bb{m, n}$, is contained within $\mathcal{M}\bb{m,n}$ (see \figurename~\ref{fig:nullspace-venn-diagram} for illustration) and is therefore measure theoretically insignificant.
		Hence, we claim that \theoremname~\ref{thm:rank-2 nullspace full} provides an almost everywhere characterization of $\mathcal{N}\bb{\mathscr{S}, 2}$.

		Of the two measure theoretically significant parts, \lemmaname~\ref{lem:rank-2 nullspace} gives an analytically simple parametrization for one part, namely $\mathcal{N}_{0}\bb{m, n}$.
		The other part, namely $\mathcal{N}_{2}\bb{m, n}$, admits a dimension recursive definition and despite our best efforts, we could not construct an analytically simple parametrization of $\mathcal{N}_{2}\bb{m, n}$ akin to \lemmaname~\ref{lem:rank-2 nullspace}.
		These results hint that development of provably correct \emph{non-randomized} coding strategies that promote signal identifiability under blind deconvolution would need to be quite sophisticated.
		In particular, the codes must disallow ambiguities arising from the recursive definition of $\mathcal{N}_{2}\bb{m, n}$.
		For provably correct \emph{simple randomized} linear coding strategies like~\cite{ahmed2012blind}, \propositionname~\ref{prop:rank-2 nullspace converse} hints that a coding redundancy of $\Theta\bb{m+n}$ is intuitively necessary to prevent bad realizations of random codes with high probability.

		\propositionname~\ref{prop:rank-2 nullspace exception set} lower bounds the dimension of the insignificant set $\mathcal{M}\bb{m,n}$ in the special case of $m = n$.
		This might be interpreted as the additional ambiguity arising from having to distinguish between equi-dimensional vectors $\vec{x},\vec{y} \in \setR^{n}$ under a bilinear operator like linear convolution; since $\bb{\vec{x}, \vec{y}}$ being a solution to the unconstrained blind deconvolution problem also implies $\bb{\vec{y}, \vec{x}}$ as a solution.
		This ambiguity does not arise when $\vec{x} \in \setR^{m}$, $\vec{y} \in \setR^{n}$ and $m \neq n$.

	\section{An Unidentifiability Result}
		\label{sec:results without coding}
		We shall use \emph{identifiability} in the sense of \definitionname~\ref{defn:identifiability}.
		After disregarding some pathological scenarios in \sectionname~\ref{sec:pathological examples} we show a strong unidentifiability result (\theoremname~\ref{thm:ae unident}) for \emph{non-sparse} blind deconvolution in \sectionname~\ref{sec:nonsparse deconv}.
		The proof strategy yields valuable insight for the sparsity constrained blind deconvolution identifiability results that we present in \partname~II of the paper~\cite{choudhary2014limitsBDsparsity}.
		Throughout this section, we assume that $\mathcal{K}$ represents the (not necessarily convex) feasible \emph{cone} in \problemname~\eqref{prob:find_xy}, \ie~$\forall \bb{\vec{x}, \vec{y}} \in \mathcal{K}$ one has $\bb{\alpha\vec{x}, \alpha\vec{y}} \in \mathcal{K}$ for every $\alpha \neq 0$.
		To prove \theoremname~\ref{thm:ae unident}, \lemmaname~\ref{lem:rank-2 nullspace} suffices and we do not need the full power of \theoremname~\ref{thm:rank-2 nullspace full}.

		\lemmaname~\ref{lem:finite quotient set} states a non-linear re-parameterization cum decomposition result to serve as a building block for constructing adversarial instances of input signals for which deconvolution fails the identifiability test.
		This is used in the proof of \theoremname~\ref{thm:ae unident} below and also to prove the results in \partname~II of the paper, and is motivated by the nonlinear parameterization of the blind deconvolution ambiguity space in \lemmaname~\ref{lem:rank-2 nullspace} and the rotational ambiguity phenomenon discussed in \sectionname~\ref{sec:examples} of the introduction.
		The astute reader would notice the symbolic connection to the transformation in \eqref{eqn:rotational transform} and the representation in \eqref{eqn:rank-2 nullspace}.

		\begin{lemma}
			\label{lem:finite quotient set}
			Let $d \geq 2$ be an arbitrary integer and $\vec{w} \in \set{\vec{w}' \in \setR^{d}}{\vec{w}'(1) \neq 0, \vec{w}'(d) \neq 0}$ be an arbitrary vector.
			The quotient set $\mathcal{Q}_{\sim}\bb{\vec{w}, d}$ defined as
			\makeatletter
				\if@twocolumn
					\begin{equation}
						\begin{split}
							\mathcal{Q}_{\sim}\bb{\vec{w}, d}
							& \triangleq	\mleft\{\bb{\vec{w}_{\ast},\gamma} \in \setR^{d-1} \times \setA \, \middle| \vphantom{\begin{bmatrix} \cos \gamma \\ \sin \gamma \end{bmatrix}} \mright.	\\
							& \qquad \quad \mleft. \vec{w} =	\begin{bmatrix}
																	\vec{w}_{\ast}	&	0	\\
																	0	&	-\vec{w}_{\ast}
																\end{bmatrix}
																\begin{bmatrix}
																	\cos \gamma	\\
																	\sin \gamma
																\end{bmatrix} \mright\},
						\end{split}
					\end{equation}
				\else
					\begin{equation}
						\mathcal{Q}_{\sim}\bb{\vec{w}, d}
						\triangleq	\set{\bb{\vec{w}_{\ast},\gamma} \in \setR^{d-1} \times \setA}
									{\vec{w} =	\begin{bmatrix}
													\vec{w}_{\ast}	&	0	\\
													0	&	-\vec{w}_{\ast}
												\end{bmatrix}
												\begin{bmatrix}
													\cos \gamma	\\
													\sin \gamma
												\end{bmatrix}},
					\end{equation}
				\fi
			\makeatother
			is finite (possibly empty) with cardinality at most $\bb{2d-2}$.
			If $d$ is an even integer then $\mathcal{Q}_{\sim}\bb{\vec{w}, d}$ is non-empty.
		\end{lemma}

		\begin{IEEEproof}
			\appendixname~\ref{sec:finite quotient set proof}.
		\end{IEEEproof}

		\subsection{Pathological Cases}
			\label{sec:pathological examples}
			Let $\vec{x}_{\ast} \in \setR^{m}$, $\vec{y}_{\ast} \in \setR^{n}$, and $\mathcal{K} = \setR^{m} \times \setR^{n}$.
			If $\vec{x}_{\ast}(m) = \vec{y}_{\ast}(1) = 0$, then $\tpose{\vec{x}_{\ast}} = \bb{\tpose{\vec{u}}, 0}$ and $\tpose{\vec{y}_{\ast}} = \bb{0, \tpose{\vec{v}}}$ for $\vec{u} = \vec{x}_{\ast}(1:m-1)$ and $\vec{v} = \vec{y}_{\ast}(2:n)$.
			\lemmaname~\ref{lem:rank-2 nullspace} and the representation in \eqref{eqn:rank-2 nullspace} imply that under the linear convolution map, $\bb{\vec{x}_{\ast}, \vec{y}_{\ast}}$ is indistinguishable from $\bb{\vec{x}',\vec{y}'}$ where $\tpose{\bb{\vec{x}'}} = \bb{0, \tpose{\vec{u}}}$ and $\tpose{\bb{\vec{y}'}} = \bb{\tpose{\vec{v}}, 0}$, since $\mat{Q} = \BB{\vec{x}_{\ast}, -\vec{x}'} \tpose{\BB{\vec{y}_{\ast}, \vec{y}'}}$ is in $\mathcal{N}\bb{\mathscr{S}, 2}$.
			As $\vec{x}_{\ast}$ and $\vec{x}'$ are linearly independent, $\bb{\vec{x}_{\ast}, \vec{y}_{\ast}}$ is unidentifiable by \definitionname~\ref{defn:identifiability}.
			Thus, for identifiability of $\bb{\vec{x}_{\ast}, \vec{y}_{\ast}}$ it is necessary that at least one of $\vec{x}_{\ast}(m)$ or $\vec{y}_{\ast}(1)$ is non-zero.
			
			Similarly, if $\vec{x}_{\ast}(1) = \vec{y}_{\ast}(n) = 0$ then invoking \lemmaname~\ref{lem:rank-2 nullspace} and representation \eqref{eqn:rank-2 nullspace} implies that $\bb{\vec{x}_{\ast}, \vec{y}_{\ast}}$ is indistinguishable (under the convolution map) from $\bb{\vec{x}', \vec{y}'}$ where $\tpose{\bb{\vec{x}'}} = \bb{\tpose{\vec{x}_{\ast}(2:m)}, 0}$ and $\tpose{\bb{\vec{y}'}} = \bb{0, \tpose{\vec{y}_{\ast}(1:n-1)}}$.
			By linear independence of $\vec{x}_{\ast}$ and $\vec{x}'$, $\bb{\vec{x}_{\ast}, \vec{y}_{\ast}}$ is unidentifiable.
			Thus, for identifiability it is necessary that at least one of $\vec{x}_{\ast}(1)$ or $\vec{y}_{\ast}(n)$ is non-zero.

			The examples above highlight that model orders $m$ and $n$ play a critical role for identifiability (and unidentifiability) for SISO systems, \ie~overestimating the model orders is fatal for blind deconvolution.
			This is a well known fact for SIMO systems~\cite{liavas1999channelorder}.

		\subsection{Non-sparse Blind Deconvolution}
			\label{sec:nonsparse deconv}
			Let $\bb{\vec{x}, \vec{y}} \in \mathcal{K} \subset \setR^{m} \times \setR^{n}$ be an arbitrary vector pair in the feasible set $\mathcal{K}$.
			To examine non-pathological examples (see \sectionname~\ref{sec:pathological examples}), we will assume that $\vec{x}(1) \neq 0$; the treatment of the other case, \viz~$\vec{y}(n) \neq 0$, is ideologically identical.

			\begin{theorem}
				\label{thm:ae unident}
				Let $m,n \geq 4$ be even integers and $\vec{x} \in \setR^{m}$ be an arbitrary vector with $\vec{x}(1) \neq 0$.
				Let the feasible set $\mathcal{K}$ be defined as
				\begin{equation}
					\mathcal{K} \triangleq \set{\vec{w} \in \setR^{m}}{\vec{w}(1) \neq 0} \times \setR^{n}.
					\label{eqn:non-sparse domain}
				\end{equation}
				Then $\bb{\vec{x}, \vec{y}} \in \mathcal{K}$ is unidentifiable almost everywhere \wrt~any measure over $\vec{y}$ that is absolutely continuous \wrt~the $n$ dimensional Lebesgue measure.
			\end{theorem}
			
			\begin{IEEEproof}
				\appendixname~\ref{sec:ae unidentifiability proof}.
			\end{IEEEproof}
			
			\theoremname~\ref{thm:ae unident} shows that unidentifiability is the norm rather than the exception.
			In fact, the number of identifiable pairs is insignificant over continuous distributions.
			
			\propositionname~\ref{prop:rank-2 nullspace converse} (and the discussion leading to it) intuitively suggests that $\Theta\bb{m}$ additional constraints on $\vec{x}$ in \problemname~\eqref{prob:find_xy} are almost necessary for identifiability \wrt~$\mathcal{K}$ (defined by \eqref{eqn:non-sparse domain}), from a heuristic DoF argument.
			If these additional constraints on $\vec{x} \in \setR^{m}$ are allowed, then it would turn \problemname~\eqref{prob:find_xy} into a semi-blind deconvolution problem.
			\theoremname~\ref{thm:ae unident} provides somewhat more concrete evidence towards the same intuition.
			Our prior work~\cite{choudhary2012sparse} discusses an example for multi-hop channel estimation, where $m-1$ additional subspace constraints are imposed on $\vec{x} \in \setR^{m}$ by system design, and this not only leads to identifiability but also to efficient and provably correct recovery algorithms.
			A philosophically similar situation is discussed in \cite{manton2003totallyblind,manton2002improvedleastsquares} using a guard interval based system for blind channel identification.

			We note that \theoremname~\ref{thm:ae unident} states a much stronger (almost everywhere \wrt~Lebesgue measure) unidentifiability result than any counterparts in the literature (\eg~\cite{kammoun2010robustness}) which only assert existence of some unidentifiable input.
			The requirement of the model orders $m$ and $n$ being even positive integers is because $\setR$ is \emph{not algebraically closed}.
			A slightly weaker version of \theoremname~\ref{thm:ae unident}, asserting the existence of a $\bb{m+n}$ dimensional set of unidentifiable signal pairs in $\mathcal{K}$ (given by \eqref{eqn:non-sparse domain}), follows readily from \theoremname~2 in \partname~II of the paper (a result on unidentifiability of blind deconvolution under canonical sparsity constraints), and indeed \textit{does not} require $m$ or $n$ to be even integers.
			This observation agrees with the absence of any conditions on the model orders in~\cite{kammoun2010robustness}.
			For completeness, we state an adapted version of \partname~II, \theoremname~2 below (which follows from setting the sparse index subsets therein to empty sets) while deferring the proof to \partname~II of the paper.
			\begin{theorem}[adapted from~\cite{choudhary2014limitsBDsparsity}]
				\label{thm:sparse unident moderate}
				Let $m,n \geq 5$ be arbitrary integers and $\mathcal{K}$ be as defined in \eqref{eqn:non-sparse domain}.
				Then there exists a set $\mathcal{G} \subseteq \mathcal{K}$ of dimension $\bb{m+n}$ such that every $\bb{\vec{x}, \vec{y}} \in \mathcal{G}$ is unidentifiable.
			\end{theorem}

			We mention in passing that the statement of \theoremname~\ref{thm:ae unident} is asymmetric \wrt~$\vec{x}$ and $\vec{y}$ since it applies to every $\vec{x}$ but not to every $\vec{y}$ (only almost every $\vec{y}$).
			This is slightly stronger than the measure theoretically symmetric version of \theoremname~\ref{thm:ae unident} that asserts unidentifiability almost everywhere \wrt~the $\bb{m+n}$ dimensional Lebesgue product measure over the pair $\bb{\vec{x}, \vec{y}}$.
			We further note that \theoremname~\ref{thm:ae unident} \emph{cannot} be strengthened to assert unidentifiability for every pair $\bb{\vec{x}, \vec{y}} \in \mathcal{K}$ as a result of a simple thought experiment.
			If such a statement were true, then reinterpreting convolution as polynomial multiplication implies that there cannot exist any polynomials over the real field that admit exactly two polynomial factors over the reals; which is clearly a false statement.

			Through the unidentifiability results in this section, we have attempted to \emph{quantify} the ill-posedness of non-sparse blind deconvolution.
			The main message is that if an application exhibits a bilinear observation model of linear convolution and no additional application specific structure can be imposed on the unknown variables then it is necessary to drastically revise the system design specifications.
			Such revision may incorporate some form of randomized precoding of the unknowns, so that the \emph{effective} bilinear operator governing the observation model looks substantially different from the convolution operator (\eg~the Gaussian random precoding used in~\cite{ahmed2012blind}).
			Alternatively, sparsity in non-canonical bases could also be helpful (\eg~Rademacher random vector signal model used in~\cite{choudhary2013bilinear}), as canonical-sparsity constraints are shown to be insufficient for identifiability in \partname~II of the paper.

	\section{Conclusions}
		\label{sec:conclusion}
		Blind deconvolution is an important non-linear inverse problem routinely encountered in signal processing applications.
		Natively, blind deconvolution is ill-posed from the viewpoint of signal identifiability and it is assumed that application specific additional constraints (like sparsity) would suffice to guarantee identifiability for this inverse problem.
		In the current work, we proved Lebesgue almost everywhere unidentifiability in non-sparse blind deconvolution for even model orders; a much stronger impossibility result than any counterparts in existing literature.
		Our approach built on the \emph{lifting} technique from optimization to reformulate blind deconvolution into a rank one matrix recovery problem, and analyzed the rank two null space of the resultant linear operator.
		While this approach is philosophically applicable to other bilinear inverse problems (like dictionary learning), it is the simplicity of the convolution operator in the lifted domain that made our analysis tractable.
		More specifically, we explicitly demonstrated a form of rotational ambiguity in bilinear functions that generates a dimension-wise large set of unidentifiable inputs.
		To establish our results, we developed a \emph{measure theoretically tight, non-linear, partially parametric and partially recursive} characterization of the rank two null space of the linear convolution map.
		Our proofs are constructive, dimensions of key sets are explicitly computed, and results hold over the real field which is not algebraically closed.
		This result is a precursor to non-randomized code design strategies for guaranteeing signal identifiability under the bilinear observation model of linear convolution.
		The design of such codes is a topic of ongoing research.
		Finally, an unanswered question that ensues from our analysis is whether the dimension-wise insignificant part $\mathcal{M}(m,n)$, of the ambiguity kernel of blind deconvolution $\mathcal{N}\bb{\mathscr{S}_{\bb{m,n}}, 2}$, is empty for $m,n \geq 3$ and $m \neq n$.

	\makeatletter
		\ifbool{@bibtex}{
			\bibliographystyle{IEEEtran}
			\bibliography{IEEEabrv,UWA,ownpub,PaperList}
		}{
			\printbibliography[heading=bibintoc]
		}
	\makeatother

	\appendices

	\section{Proof of \propositionname~\ref{prop:rank-2 nullspace converse}}
		\label{sec:nullspace converse proof}
		The proof involves contradiction by an explicit dimension recursive construction.
		Let $m,n \geq 3$ be arbitrary integers and let $\mathscr{S}'\fcolon \setR^{(m-1) \times (n-1)} \to \setR^{m+n-3}$ denote the lifted linear convolution map in one lower dimension than $\mathscr{S}\bb{\cdot}$ \wrt~both rows and columns.
		Further, let $\vec{u}_{\ast} \in \setR^{m-2} \setminus \cc{\vec{0}}$, $\vec{v}_{\ast} \in \setR^{n-2} \setminus \cc{\vec{0}}$ be arbitrary vectors and define $\mat{X} \in \setR^{(m-1) \times (n-1)}$ as
		\begin{equation}
			\mat{X} =	\begin{bmatrix}
							\vec{u}_{\ast}	&	0	\\
							0	&	-\vec{u}_{\ast}
						\end{bmatrix}
						\begin{bmatrix}
							0	&	\tpose{\vec{v}_{\ast}}	\\
							\tpose{\vec{v}_{\ast}}	&	0
						\end{bmatrix}.
			\label{eqn:nullspace element}
		\end{equation}
		Since $(m-1) \geq 2$ and $(n-1) \geq 2$, using \lemmaname~\ref{lem:rank-2 nullspace} on $\mat{X}$ implies $\mat{X} \in \mathcal{N}\bb{\mathscr{S}', 2}$.
		For any invertible matrix $\mat{A} \in \setR^{2 \times 2}$, let us designate
		\begin{alignat}{2}
			\begin{bmatrix}
				\vec{u}_{\ast}	&	0	\\
				0	&	-\vec{u}_{\ast}
			\end{bmatrix}	\mat{A}
			& =	\begin{bmatrix}
					\vec{u}_{1}	&	\vec{u}_{2}
				\end{bmatrix},%
			& \quad \inv{\mat{A}}	\begin{bmatrix}
										0	&	\tpose{\vec{v}_{\ast}}	\\
										\tpose{\vec{v}_{\ast}}	&	0
									\end{bmatrix}
			& =	\begin{bmatrix}
					\tpose{\vec{v}_{1}}	\\
					\tpose{\vec{v}_{2}}
				\end{bmatrix}
		\end{alignat}
		for some $\vec{u}_{1}, \vec{u}_{2} \in \setR^{m-1}$ and $\vec{v}_{1}, \vec{v}_{2} \in \setR^{n-1}$.
		Clearly, $\mat{X} = \vec{u}_{1} \tpose{\vec{v}_{1}} + \vec{u}_{2} \tpose{\vec{v}_{2}}$.
		Furthermore, the vectors $\cc{\vec{u}_{1}, \vec{u}_{2}}$ are linearly independent, owing to the invertibility of $\mat{A}$.
		Consider the matrix $\mat{Y} \in \setR^{m \times n}$ constructed as
		\begin{equation}
			\mat{Y}	=	\begin{bmatrix}
							\vec{u}_{1}	&	0	\\
							0	&	\vec{u}_{2}
						\end{bmatrix}
						\begin{bmatrix}
							0	&	\tpose{\vec{v}_{1}}	\\
							\tpose{\vec{v}_{2}}	&	0
						\end{bmatrix}
					=	\underbrace{\begin{bmatrix}
										\vec{0}	&	\vec{u}_{1} \tpose{\vec{v}_{1}}	\\
										0	&	\tpose{\vec{0}}
									\end{bmatrix}}_{\mat{Y}_{1}}
					+	\underbrace{\begin{bmatrix}
										\tpose{\vec{0}}	&	0	\\
										\vec{u}_{2} \tpose{\vec{v}_{2}}	&	\vec{0}
									\end{bmatrix}}_{\mat{Y}_{2}}.
			\label{eqn:one higher dimension}
		\end{equation}
		From \eqref{eqn:block shifted construction} and \eqref{eqn:equality under convolution} in \sectionname~\ref{sec:anti-diagonal sum interpretation}, $\mathscr{S}\bb{\mat{Y}_{1}}$ does not change if elements of $\mat{Y}_{1}$ are shifted down by one unit along the anti-diagonals.
		This implies
		\begin{equation}
			\mathscr{S}\bb{\mat{Y}_{j}} =	\begin{bmatrix}
												0	\\
												\mathscr{S}'\bb{\vec{u}_{j} \tpose{\vec{v}_{j}}}	\\
												0
											\end{bmatrix}
		\end{equation}
		for $j = 1,2$ and thus,
		\begin{equation}
			\mathscr{S}\bb{\mat{Y}} =	\begin{bmatrix}
											0	\\
											\mathscr{S}'\bb{\vec{u}_{1} \tpose{\vec{v}_{1}} + \vec{u}_{2} \tpose{\vec{v}_{2}}}	\\
											0
										\end{bmatrix}
			=	\begin{bmatrix}
					0	\\
					\mathscr{S}'\bb{\mat{X}}	\\
					0
				\end{bmatrix}
			=	\vec{0}
			\label{eqn:zero result}
		\end{equation}
		using \eqref{eqn:one higher dimension} and $\vec{u}_{1} \tpose{\vec{v}_{1}} + \vec{u}_{2} \tpose{\vec{v}_{2}} = \mat{X} \in \mathcal{N}\bb{\mathscr{S}', 2}$.
		Therefore, $\mat{Y} \in \mathcal{N}\bb{\mathscr{S}, 2}$ since \eqref{eqn:one higher dimension} implies $\rank{\mat{Y}} \leq 2$, but $\mat{Y}$ as represented in \eqref{eqn:one higher dimension} is not convertible to the form in \eqref{eqn:rank-2 nullspace} because of linear independence of $\vec{u}_{1}$ and $\vec{u}_{2}$.

		To finish the proof, we need to show that the parametric family of matrices $\mat{Y}$ as defined by \eqref{eqn:one higher dimension} is of dimension $(m+n-3)$.
		We denote this set by $\mathcal{M}$.
		Consider $\twonorm{\vec{u}_{\ast}} = \twonorm{\vec{v}_{\ast}} = 1$ so that $\vec{u}_{\ast} \in \setR^{m-2}$ is $(m-3)$ dimensional and $\vec{v}_{\ast} \in \setR^{n-2}$ is $(n-3)$ dimensional.
		$\mat{A}$ represents two independently chosen directions of linear combination to generate $\vec{u}_{1}$ and $\vec{u}_{2}$, so $\mat{A}$ is two dimensional.
		Finally, given our calculations, we need to add one dimension for a scalar multiplicative parameter to account for all matrices $\mat{Y} \in \mathcal{M}$.
		Thus, $\mathcal{M}$ is $(m-3) + (n-3) + 2 + 1 = (m+n-3)$ dimensional.

	\section{Proof of \theoremname~\ref{thm:rank-2 nullspace full}}
		\label{sec:full null space proof}
		Throughout this proof, we shall let $\phi_{j}$ denote the $j^{\thp}$ coordinate projection operator, \ie~$\phi_{j}\bb{\vec{x}} = \vec{x}(j)$.

		\subsection{Proof of \partname~\ref{itm:m1 n1}}
			\label{sec:full null space proof part 1}
			We argue that $\mathcal{N}\bb{\mathscr{S}_{\bb{1,n}}, 2} = \cc{\mat{0}}$.
			The proof for $\mathcal{N}\bb{\mathscr{S}_{\bb{m,1}}, 2} = \cc{\mat{0}}$ is similar.
			Since linear convolution with a scalar is equivalent to multiplication by the same scalar, by definition of $\mathscr{S}_{\bb{1,n}} \fcolon \setR^{n} \to \setR^{n}$, the lifted operator $\mathscr{S}_{\bb{1,n}}\bb{\cdot}$ is an identity operator on $\setR^{n}$.
			This immediately implies $\mathcal{N}\bb{\mathscr{S}_{\bb{1,n}}, 2} = \cc{\mat{0}}$ since the identity operator has a trivial null space.

		\subsection{Proof of \partname~\ref{itm:m2 n2}}
			\label{sec:full null space proof part 2}
			Let $n \geq 2$ be an arbitrary integer.
			We show that $\mathcal{N}\bb{\mathscr{S}_{\bb{2,n}}, 2} = \mathcal{N}_{0}\bb{2,n}$.
			The proof of $\mathcal{N}\bb{\mathscr{S}_{\bb{m,2}}, 2} = \mathcal{N}_{0}\bb{m, 2}$ for $m \geq 2$ is similar.
			Setting $m = 2$ in \lemmaname~\ref{lem:rank-2 nullspace}, we have $\mathcal{N}_{0}\bb{2, n} \subseteq \mathcal{N}\bb{\mathscr{S}_{\bb{2,n}}, 2}$.
			To complete the proof, we need to show $\mathcal{N}\bb{\mathscr{S}_{\bb{2,n}}, 2} \subseteq \mathcal{N}_{0}\bb{2, n}$.
			Let $\mat{X} \in \mathcal{N}\bb{\mathscr{S}_{\bb{2,n}}, 2} \subset \setR^{2 \times n}$.
			We have $\phi_{1} \circ \mathscr{S}_{\bb{2,n}}\bb{\mat{X}} = 0$ and by definition of $\mathscr{S}_{\bb{2,n}}\bb{\cdot}$, we have $\mat{X}\bb{1,1} = \phi_{1} \circ \mathscr{S}_{\bb{2,n}}\bb{\mat{X}}$ (see \figurename~\ref{fig:anti-diagonal sum} for illustration), implying that $\mat{X}\bb{1,1} = 0$.
			By a similar argument, $\mat{X}\bb{2,n} = \phi_{n+1} \circ \mathscr{S}_{\bb{2,n}}\bb{\mat{X}} = 0$.
			By definition of $\mathscr{S}_{\bb{2,n}}\bb{\cdot}$ we have,
			\begin{equation}
				\mathscr{S}_{\bb{2,n}}\bb{\mat{X}} =	\begin{bmatrix}
															0	\\
															\mat{X}(1,2:n) + \mat{X}(2,1:n-1)	\\
															0
														\end{bmatrix}
													=	\vec{0}
			\end{equation}
			implying that $\mat{X}(1,2:n) = -\mat{X}(2,1:n-1)$.
			Letting $\tpose{\vec{v}} = \mat{X}(1,2:n)$ we get
			\begin{equation}
				\mat{X} =	\begin{bmatrix}
								1	&	0	\\
								0	&	-1
							\end{bmatrix}
							\begin{bmatrix}
								0	&	\tpose{\vec{v}}	\\
								\tpose{\vec{v}}	&	0
							\end{bmatrix}
						\in \mathcal{N}_{0}\bb{2, n}.
			\end{equation}
			Since $\mat{X} \in \mathcal{N}\bb{\mathscr{S}_{\bb{2,n}}, 2}$ is arbitrary, we have $\mathcal{N}\bb{\mathscr{S}_{\bb{2,n}}, 2} \subseteq \mathcal{N}_{0}\bb{2, n}$.

		\subsection{Proof of \partname~\ref{itm:m n subset}}
			\label{sec:full null space proof part 3}
			Let $m,n \geq 3$ be arbitrary integers.
			We'll show that $\mathcal{N}_{0}\bb{m, n} \bigcup \mathcal{N}_{2}\bb{m, n} \subseteq \mathcal{N}\bb{\mathscr{S}_{\bb{m,n}}, 2}$.
			From \lemmaname~\ref{lem:rank-2 nullspace}, we have $\mathcal{N}_{0}\bb{m, n} \subseteq \mathcal{N}\bb{\mathscr{S}_{\bb{m,n}}, 2}$.
			It remains to show that $\mathcal{N}_{2}\bb{m, n} \subseteq \mathcal{N}\bb{\mathscr{S}_{\bb{m,n}}, 2}$.
			To this end, we borrow constructions from the proof of \propositionname~\ref{prop:rank-2 nullspace converse}.
			Let $\mat{X} = \vec{u}_{1}\tpose{\vec{v}_{1}} + \vec{u}_{2}\tpose{\vec{v}_{2}} \in \mathcal{N}\bb{\mathscr{S}_{\bb{m-1,n-1}}, 2} \setminus \cc{\mat{0}}$ for some $\vec{u}_{1}, \vec{u}_{2} \in \setR^{m-1}$ and $\vec{v}_{1}, \vec{v}_{2} \in \setR^{n-1}$.
			We construct $\mat{Y} \in \mathcal{N}_{2}\bb{m,n} \subset \setR^{m \times n}$ as in \eqref{eqn:one higher dimension} and following the chain of arguments \eqref{eqn:one higher dimension} $\rightarrow$ \eqref{eqn:zero result} using the anti-diagonal sum interpretation of \sectionname~\ref{sec:anti-diagonal sum interpretation}, we have
			\makeatletter
				\if@twocolumn
					\begin{equation}
						\begin{split}
							\mathscr{S}_{\bb{m,n}}\bb{\mat{Y}} & =	\begin{bmatrix}
																		0	\\
																		\mathscr{S}_{\bb{m-1,n-1}}\bb{\vec{u}_{1} \tpose{\vec{v}_{1}} + \vec{u}_{2} \tpose{\vec{v}_{2}}}	\\
																		0
																	\end{bmatrix}	\\
							& =	\begin{bmatrix}
									0	\\
									\mathscr{S}_{\bb{m-1,n-1}}\bb{\mat{X}}	\\
									0
								\end{bmatrix}
							=	\vec{0}
						\end{split}
					\end{equation}
				\else
					\begin{equation}
						\mathscr{S}_{\bb{m,n}}\bb{\mat{Y}} =	\begin{bmatrix}
																	0	\\
																	\mathscr{S}_{\bb{m-1,n-1}}\bb{\vec{u}_{1} \tpose{\vec{v}_{1}} + \vec{u}_{2} \tpose{\vec{v}_{2}}}	\\
																	0
																\end{bmatrix}
															=	\begin{bmatrix}
																	0	\\
																	\mathscr{S}_{\bb{m-1,n-1}}\bb{\mat{X}}	\\
																	0
																\end{bmatrix}
															=	\vec{0}
					\end{equation}
				\fi
			\makeatother
			since $\mat{X} \in \mathcal{N}\bb{\mathscr{S}_{\bb{m-1,n-1}}, 2}$.
			Further, \eqref{eqn:one higher dimension} implies that $\rank{\mat{Y}} \leq 2$ and therefore $\mat{Y} \in \mathcal{N}\bb{\mathscr{S}_{\bb{m,n}}, 2}$.
			Since $\mat{X} \in \mathcal{N}\bb{\mathscr{S}_{\bb{m-1,n-1}}, 2} \setminus \cc{\mat{0}}$ is arbitrary, we have $\mathcal{N}_{2}\bb{m, n} \subseteq \mathcal{N}\bb{\mathscr{S}_{\bb{m,n}}, 2}$.

		\subsection{Proof of \partname~\ref{itm:m n restricted}}
			\label{sec:full null space proof part 4}
			Let $m,n \geq 3$ be arbitrary integers.
			From the preceding part, we have $\mathcal{N}_{0}\bb{m, n} \bigcup \mathcal{N}_{2}\bb{m, n} \subseteq \mathcal{N}\bb{\mathscr{S}_{\bb{m,n}}, 2}$ implying that $\bb[\big]{\mathcal{N}_{0}\bb{m, n} \bigcup \mathcal{N}_{2}\bb{m, n}} \setminus \mathcal{M}(m,n) \subseteq \mathcal{N}\bb{\mathscr{S}_{\bb{m,n}}, 2} \setminus \mathcal{M}(m,n)$.
			It remains to show that $\mathcal{N}\bb{\mathscr{S}_{\bb{m,n}}, 2} \setminus \mathcal{M}(m,n) \subseteq \bb[\big]{\mathcal{N}_{0}\bb{m, n} \bigcup \mathcal{N}_{2}\bb{m, n}} \setminus \mathcal{M}(m,n)$.
			Let $\mat{X} \in \mathcal{N}\bb{\mathscr{S}_{\bb{m,n}}, 2} \setminus \mathcal{M}(m,n) \subset \setR^{m \times n}$.
			We have $\phi_{1} \circ \mathscr{S}_{\bb{m,n}}\bb{\mat{X}} = 0$ and by definition of $\mathscr{S}_{\bb{m,n}}\bb{\cdot}$, we have $\mat{X}\bb{1,1} = \phi_{1} \circ \mathscr{S}_{\bb{m,n}}\bb{\mat{X}}$ (see \figurename~\ref{fig:anti-diagonal sum} for illustration), implying that $\mat{X}\bb{1,1} = 0$.
			By a similar argument, $\mat{X}(m,n) = \phi_{m+n-1} \circ \mathscr{S}_{\bb{m,n}}\bb{\mat{X}} = 0$.
	
			\emph{Our proof strategy will be to show that $\mat{X}$ always admits a factorization like \eqref{eqn:one higher dimension} and then deduce that this necessarily implies our result.}
			Since $\mat{X} \not \in \mathcal{M}\bb{m,n}$, either $\mat{X}(m,1) \neq 0$ or $\mat{X}(1,n) \neq 0$ is true.
			Let us first assume that $\mat{X}(m,1) \neq 0$ implying that $\mat{X}(:,1) \neq \vec{0}$.
			We have two further possibilities, \viz~$\mat{X}(1,:) \neq \tpose{\vec{0}}$ and $\mat{X}(1,:) = \tpose{\vec{0}}$.
	
			\subsubsection{First Case}
				\label{itm:not zero}
				Suppose that $\mat{X}(1,:) \neq \tpose{\vec{0}}$.
				We set $\vec{u}_{2} = \mat{X}\bb{2:m,1} \in \setR^{m-1}$ so that $\BB{0, \tpose{\vec{u}_{2}}} = \tpose{\mat{X}\bb{:,1}}$.
				Since $\mathcal{N}\bb{\mathscr{S}_{\bb{m,n}}, 1} = \cc{\mat{0}}$, we have $\rank{\mat{X}} = 2$ and since $\mat{X}(:,1) \neq \vec{0}$ we conclude that $\exists j_{0} \in \cc{2,3,\dots,n}$ such that $\mat{X}\bb{:,1}$ and $\mat{X}\bb{:,j_{0}}$ are linearly independent.
				We choose $\vec{u}_{1} \in \setR^{m-1}$ as follows.
				If $\mat{X}\bb{m,j_{0}} = 0$ (\eg~$\mat{X}\bb{m,n} = 0$ for $j_{0} = n$), then we set $\vec{u}_{1} = \mat{X}\bb{1:m-1,j_{0}}$, else we set
				\begin{equation}
					\vec{u}_{1} = \frac{\mat{X}\bb{m,1}}{\mat{X}\bb{m,j_{0}}} \mat{X}\bb{1:m-1,j_{0}} - \mat{X}\bb{1:m-1,1}.
				\end{equation}
				We claim that the vectors $\BB{\tpose{\vec{u}_{1}}, 0}$ and $\BB{0, \tpose{\vec{u}_{2}}}$ are linearly independent and span $\mathcal{C}\bb{\mat{X}}$, which would imply that $\mat{X}$ admits a factorization of the form
				\begin{equation}
					\mat{X} =	\begin{bmatrix}
									\vec{u}_{1}	&	0	\\
									0	&	\vec{u}_{2}
								\end{bmatrix}
								\begin{bmatrix}
									\tpose{\vec{v}}	\\
									\tpose{\vec{v}_{\ast}}
								\end{bmatrix}
					\label{eqn:interim factorization}
				\end{equation}
				for some vectors $\vec{v}, \vec{v}_{\ast} \in \setR^{n}$.
				Indeed, if $\mat{X}\bb{m,j_{0}} = 0$ then $\BB{\tpose{\vec{u}_{1}}, 0} = \tpose{\mat{X}\bb{:,j_{0}}}$ and $\tpose{\mat{X}\bb{:,1}} = \BB{0, \tpose{\vec{u}_{2}}}$ are linearly independent and span $\mathcal{C}\bb{\mat{X}}$ since $\rank{\mat{X}} = 2$.
				In case $\mat{X}\bb{m,j_{0}} \neq 0$, setting $\alpha = \mat{X}\bb{m,1}/\mat{X}\bb{m,j_{0}}$ we observe that
				\begin{equation}
					\begin{bmatrix}
						\vec{u}_{1}	&	0	\\
						0	&	\vec{u}_{2}
					\end{bmatrix}
					=	\begin{bmatrix}
							\mat{X}\bb{:,1}	&	\mat{X}\bb{:,j_{0}}
						\end{bmatrix}
						\begin{bmatrix}
								-1	&	1	\\
							\alpha	&	0
						\end{bmatrix}.
					\label{eqn:column space representation}
				\end{equation}
				Since $\mat{X}(m,1) \neq 0$ and $\mat{X}(m,j_{0}) \neq 0$, $\alpha \in \setR \setminus \cc{\mat{0}}$ and \eqref{eqn:column space representation} imply that $\BB{\tpose{\vec{u}_{1}}, 0}$ and $\BB{0, \tpose{\vec{u}_{2}}}$ are linearly independent and span $\mathcal{C}\bb{\mat{X}}$ owing to the linear independence of $\mat{X}(:,1)$ and $\mat{X}(:,j_{0})$ and $\rank{\mat{X}}$ being equal to two.
				Thus, \eqref{eqn:interim factorization} holds implying that $\vec{u}_{1}(1) \vec{v}(1) = \mat{X}(1,1) = 0$ and $\vec{u}_{2}(m-1) \vec{v}_{\ast}(n) = \mat{X}(m,n) = 0$.
				We have $\vec{u}_{2}(m-1) = \mat{X}(m,1) \neq 0$ by construction so $\vec{v}_{\ast}(n) = 0$.
				Further, if $\vec{u}_{1}(1) = 0$ were true then \eqref{eqn:interim factorization} would imply $\mat{X}(1,:) = \tpose{\vec{0}}$ contradicting our assumption of $\mat{X}(1,:) \neq \tpose{\vec{0}}$.
				Thus, $\vec{u}_{1}(1) \neq 0$ which implies $\vec{v}(1) = \mat{X}(1,1)/\vec{u}_{1}(1) = 0$.
				Letting $\vec{v}_{1} = \vec{v}(2:n)$ and $\vec{v}_{2} = \vec{v}_{\ast}(1:n-1)$, \eqref{eqn:interim factorization} implies
				\begin{equation}
					\mat{X} =	\begin{bmatrix}
									\vec{u}_{1}	&	0	\\
									0	&	\vec{u}_{2}
								\end{bmatrix}
								\begin{bmatrix}
									0	&	\tpose{\vec{v}_{1}}	\\
									\tpose{\vec{v}_{2}}	&	0
								\end{bmatrix}
							=	\underbrace{\begin{bmatrix}
												\vec{0}	&	\vec{u}_{1} \tpose{\vec{v}_{1}}	\\
												0	&	\tpose{\vec{0}}
											\end{bmatrix}}_{\mat{X}_{1}}
							+	\underbrace{\begin{bmatrix}
											\tpose{\vec{0}}	&	0	\\
											\vec{u}_{2} \tpose{\vec{v}_{2}}	&	\vec{0}
											\end{bmatrix}}_{\mat{X}_{2}}.
					\label{eqn:null matrix factoring}
				\end{equation}
				Following the same chain of arguments as \eqref{eqn:one higher dimension} $\rightarrow$ \eqref{eqn:zero result} with the anti-diagonal sum interpretation, \eqref{eqn:null matrix factoring} implies
				\makeatletter
					\if@twocolumn
						\begin{equation}
							\begin{split}
								\MoveEqLeft	\begin{bmatrix}
												0	\\
												\mathscr{S}_{\bb{m-1,n-1}}\bb{\vec{u}_{1} \tpose{\vec{v}_{1}} + \vec{u}_{2} \tpose{\vec{v}_{2}}}	\\
												0
											\end{bmatrix}	\\
								& =	\mathscr{S}_{\bb{m,n}}\bb{\mat{X}_{1}} + \mathscr{S}_{\bb{m,n}}\bb{\mat{X}_{2}}
								=	\vec{0}
							\end{split}
						\end{equation}
					\else
						\begin{equation}
							\begin{bmatrix}
								0	\\
								\mathscr{S}_{\bb{m-1,n-1}}\bb{\vec{u}_{1} \tpose{\vec{v}_{1}} + \vec{u}_{2} \tpose{\vec{v}_{2}}}	\\
								0
							\end{bmatrix}
							=	\mathscr{S}_{\bb{m,n}}\bb{\mat{X}_{1}} + \mathscr{S}_{\bb{m,n}}\bb{\mat{X}_{2}}
							=	\vec{0}
						\end{equation}
					\fi
				\makeatother
				since $\mat{X} \in \mathcal{N}\bb{\mathscr{S}_{\bb{m,n}}, 2}$.
				Therefore, $\vec{u}_{1} \tpose{\vec{v}_{1}} + \vec{u}_{2} \tpose{\vec{v}_{2}} \in \mathcal{N}\bb{\mathscr{S}_{\bb{m-1,n-1}}, 2}$.
				If $\vec{u}_{1} \tpose{\vec{v}_{1}} + \vec{u}_{2} \tpose{\vec{v}_{2}} \neq \mat{0}$ then \eqref{eqn:null matrix factoring} and \eqref{eqn:rank-2 nullspace part 2} imply that $\mat{X} \in \mathcal{N}_{2}\bb{m, n}$.
				Instead, if $\vec{u}_{1} \tpose{\vec{v}_{1}} + \vec{u}_{2} \tpose{\vec{v}_{2}} = \mat{0}$ then $\vec{u}_{2} \tpose{\vec{v}_{2}} = -\vec{u}_{1} \tpose{\vec{v}_{1}}$ and \eqref{eqn:null matrix factoring} implies
				\makeatletter
					\if@twocolumn
						\begin{equation}
							\begin{split}
								\mat{X}	& = \mat{X}_{1} + \mat{X}_{2}
										=	\begin{bmatrix}
												\vec{0}	&	\vec{u}_{1} \tpose{\vec{v}_{1}}	\\
												0	&	\tpose{\vec{0}}
											\end{bmatrix}
										+	\begin{bmatrix}
												\tpose{\vec{0}}	&	0	\\
												-\vec{u}_{1} \tpose{\vec{v}_{1}}	&	\vec{0}
											\end{bmatrix}	\\
										& =	\begin{bmatrix}
												\vec{u}_{1}	&	0	\\
												0	&	-\vec{u}_{1}
											\end{bmatrix}
											\begin{bmatrix}
												0	&	\tpose{\vec{v}_{1}}	\\
												\tpose{\vec{v}_{1}}	&	0
											\end{bmatrix}
							\end{split}
						\end{equation}
					\else
						\begin{equation}
							\mat{X}	= \mat{X}_{1} + \mat{X}_{2}
									=	\begin{bmatrix}
											\vec{0}	&	\vec{u}_{1} \tpose{\vec{v}_{1}}	\\
											0	&	\tpose{\vec{0}}
										\end{bmatrix}
									+	\begin{bmatrix}
											\tpose{\vec{0}}	&	0	\\
											-\vec{u}_{1} \tpose{\vec{v}_{1}}	&	\vec{0}
										\end{bmatrix}
									=	\begin{bmatrix}
											\vec{u}_{1}	&	0	\\
											0	&	-\vec{u}_{1}
										\end{bmatrix}
										\begin{bmatrix}
											0	&	\tpose{\vec{v}_{1}}	\\
											\tpose{\vec{v}_{1}}	&	0
										\end{bmatrix}
						\end{equation}
					\fi
				\makeatother
				and thus $\mat{X} \in \mathcal{N}_{0}\bb{m, n}$ by \eqref{eqn:rank-2 nullspace part 1}.
				Since $\mat{X} \not \in \mathcal{M}\bb{m,n}$ by assumption, we get $\mat{X} \in \bb[\big]{\mathcal{N}_{0}\bb{m, n} \bigcup \mathcal{N}_{2}\bb{m, n}} \setminus \mathcal{M}(m,n)$.
	
			\subsubsection{Second Case}
				Now suppose that $\mat{X}(1,:) = \tpose{\vec{0}}$ so that
				\begin{equation}
					\mat{X} =	\begin{bmatrix}
									\tpose{\vec{0}}	\\
									\mat{Y}
								\end{bmatrix}
					\label{eqn:interim representation}
				\end{equation}
				for some matrix $\mat{Y} \in \setR^{\bb{m-1} \times n}$.
				We will use mathematical induction on $m$ to settle this case.
	
				\textit{Induction Step:}	Let $\mathcal{N}\bb{\mathscr{S}_{\bb{m',n}}, 2} \setminus \mathcal{M}(m',n) \subseteq \bb[\big]{\mathcal{N}_{0}\bb{m', n} \bigcup \mathcal{N}_{2}\bb{m', n}} \setminus \mathcal{M}(m',n)$ be true for $m' = m-1 \geq 3$.
				Since $\mathscr{S}_{\bb{m,n}}\bb{\cdot}$ sums elements along the anti-diagonals (see \figurename~\ref{fig:anti-diagonal sum} for illustration) and $\mat{X} \in \mathcal{N}\bb{\mathscr{S}_{\bb{m,n}}, 2}$, we have
				\begin{equation}
					\begin{bmatrix}
						0	\\
						\mathscr{S}_{\bb{m',n}}\bb{\mat{Y}}
					\end{bmatrix}
					=	\mathscr{S}_{\bb{m,n}}\bb{\mat{X}}
					=	\vec{0}
				\end{equation}
				implying that $\mat{Y} \in \mathcal{N}\bb{\mathscr{S}_{\bb{m',n}}, 2}$.
				Further, $\mat{X}(m,1) \neq 0$ by assumption and $\mat{Y}(m',1) = \mat{X}(m,1)$ by \eqref{eqn:interim representation} implying that $\mat{Y} \not \in \mathcal{M}(m',n)$ by definition.
				Thus, $\mat{Y} \in \mathcal{N}\bb{\mathscr{S}_{\bb{m',n}}, 2} \setminus \mathcal{M}(m',n)$ and by the induction hypothesis either $\mat{Y} \in \mathcal{N}_{2}\bb{m', n}$ or $\mat{Y} \in \mathcal{N}_{0}\bb{m', n}$.
				We separate the analysis for each of these cases below.
	
				\begin{enumerate}
					\item	Suppose that $\mat{Y} \in \mathcal{N}_{2}\bb{m', n}$.
					Then there exist vectors $\vec{u}_{1}, \vec{u}_{2} \in \setR^{m'-1}$ and $\vec{v}_{1}, \vec{v}_{2} \in \setR^{n-1}$ such that
					\begin{equation}
						\mat{Y} =	\begin{bmatrix}
										\vec{u}_{1}	&	0	\\
										0	&	\vec{u}_{2}
									\end{bmatrix}
									\begin{bmatrix}
										0	&	\tpose{\vec{v}_{1}}	\\
										\tpose{\vec{v}_{2}}	&	0
									\end{bmatrix}
					\end{equation}
					and $\vec{u}_{1} \tpose{\vec{v}_{1}} + \vec{u}_{2} \tpose{\vec{v}_{2}} \in \mathcal{N}\bb{\mathscr{S}_{\bb{m'-1,n-1}}, 2} \setminus \cc{\mat{0}}$.
					Now using \eqref{eqn:interim representation} we get
					\begin{equation}
						\mat{X} =	\begin{bmatrix}
										0	&	0	\\
										\vec{u}_{1}	&	0	\\
										0	&	\vec{u}_{2}
									\end{bmatrix}
									\begin{bmatrix}
										0	&	\tpose{\vec{v}_{1}}	\\
										\tpose{\vec{v}_{2}}	&	0
									\end{bmatrix}
								=	\begin{bmatrix}
										\vec{u}_{3}	&	0	\\
										0	&	\vec{u}_{4}
									\end{bmatrix}
									\begin{bmatrix}
										0	&	\tpose{\vec{v}_{1}}	\\
										\tpose{\vec{v}_{2}}	&	0
									\end{bmatrix}
					\end{equation}
					where we have set $\tpose{\vec{u}_{3}} = \BB{0, \tpose{\vec{u}_{1}}}$ and $\tpose{\vec{u}_{4}} = \BB{0, \tpose{\vec{u}_{2}}}$.
					Clearly, $\vec{u}_{3}, \vec{u}_{4} \in \setR^{m-1}$.
					To conclude that $\mat{X} \in \mathcal{N}_{2}\bb{m,n}$ we need to show that $\vec{u}_{3}\tpose{\vec{v}_{1}} + \vec{u}_{4}\tpose{\vec{v}_{2}} \in \mathcal{N}\bb{\mathscr{S}_{\bb{m-1,n-1}}, 2} \setminus \cc{\mat{0}}$.
					We have,
					\begin{equation}
						\vec{u}_{3}\tpose{\vec{v}_{1}} + \vec{u}_{4}\tpose{\vec{v}_{2}}
						=	\begin{bmatrix}
								0	\\
								\vec{u}_{1}
							\end{bmatrix}	\tpose{\vec{v}_{1}}
						+	\begin{bmatrix}
								0	\\
								\vec{u}_{2}
							\end{bmatrix}	\tpose{\vec{v}_{2}}
						=	\begin{bmatrix}
								\tpose{\vec{0}}	\\
								\vec{u}_{1}\tpose{\vec{v}_{1}} + \vec{u}_{2}\tpose{\vec{v}_{2}}
							\end{bmatrix}
						\label{eqn:one dimension lift}
					\end{equation}
					and therefore
					\makeatletter
						\if@twocolumn
							\begin{equation}
								\begin{split}
									\MoveEqLeft[1] \mathscr{S}_{\bb{m-1,n-1}}\bb{\vec{u}_{3}\tpose{\vec{v}_{1}} + \vec{u}_{4}\tpose{\vec{v}_{2}}}	\\
									& =	\begin{bmatrix}
											0	\\
											\mathscr{S}_{\bb{m-2,n-1}}\bb{\vec{u}_{1}\tpose{\vec{v}_{1}} + \vec{u}_{2}\tpose{\vec{v}_{2}}}
										\end{bmatrix}
									=	\vec{0}
								\end{split}
								\label{eqn:null space verification}
							\end{equation}
						\else
							\begin{equation}
								\mathscr{S}_{\bb{m-1,n-1}}\bb{\vec{u}_{3}\tpose{\vec{v}_{1}} + \vec{u}_{4}\tpose{\vec{v}_{2}}}
								=	\begin{bmatrix}
										0	\\
										\mathscr{S}_{\bb{m-2,n-1}}\bb{\vec{u}_{1}\tpose{\vec{v}_{1}} + \vec{u}_{2}\tpose{\vec{v}_{2}}}
									\end{bmatrix}
								=	\vec{0}
								\label{eqn:null space verification}
							\end{equation}
						\fi
					\makeatother
					where the first equality is because $\mathscr{S}_{\bb{m-1,n-1}}\bb{\cdot}$ sums elements along the anti-diagonals and the second equality follows from the fact that $\vec{u}_{1} \tpose{\vec{v}_{1}} + \vec{u}_{2} \tpose{\vec{v}_{2}} \in \mathcal{N}\bb{\mathscr{S}_{\bb{m'-1,n-1}}, 2} \setminus \cc{\mat{0}}$ and $m' = m-1$.
					Since $\vec{u}_{1} \tpose{\vec{v}_{1}} + \vec{u}_{2} \tpose{\vec{v}_{2}} \neq \mat{0}$, \eqref{eqn:one dimension lift} implies that $\vec{u}_{3}\tpose{\vec{v}_{1}} + \vec{u}_{4}\tpose{\vec{v}_{2}} \neq \mat{0}$.
					Further, $\rank{\vec{u}_{3}\tpose{\vec{v}_{1}} + \vec{u}_{4}\tpose{\vec{v}_{2}}} \leq 2$ and thus, \eqref{eqn:null space verification} implies that $\vec{u}_{3}\tpose{\vec{v}_{1}} + \vec{u}_{4}\tpose{\vec{v}_{2}} \in \mathcal{N}\bb{\mathscr{S}_{\bb{m-1,n-1}}, 2} \setminus \cc{\mat{0}}$ and hence $\mat{X} \in \mathcal{N}_{2}\bb{m,n}$.
					By assumption, $\mat{X} \not \in \mathcal{M}\bb{m,n}$ and therefore $\mat{X} \in \mathcal{N}_{2}\bb{m,n} \setminus \mathcal{M}\bb{m,n}$.
	
					\item	Assume that $\mat{Y} \in \mathcal{N}_{0}\bb{m', n}$.
					Then there exist vectors $\vec{u} \in \setR^{m'-1}$ and $\vec{v} \in \setR^{n-1}$ such that
					\begin{equation}
						\mat{Y} =	\begin{bmatrix}
										\vec{u}	&	0	\\
										0	&	-\vec{u}
									\end{bmatrix}
									\begin{bmatrix}
										0	&	\tpose{\vec{v}}	\\
										\tpose{\vec{v}}	&	0
									\end{bmatrix}
					\end{equation}
					and \eqref{eqn:interim representation} gives
					\begin{equation}
						\mat{X} =	\begin{bmatrix}
										0	&	0	\\
										\vec{u}	&	0	\\
										0	&	-\vec{u}
									\end{bmatrix}
									\begin{bmatrix}
										0	&	\tpose{\vec{v}}	\\
										\tpose{\vec{v}}	&	0
									\end{bmatrix}
								=	\begin{bmatrix}
										\vec{u}_{\ast}	&	0	\\
										0	&	-\vec{u}_{\ast}
									\end{bmatrix}
									\begin{bmatrix}
										0	&	\tpose{\vec{v}}	\\
										\tpose{\vec{v}}	&	0
									\end{bmatrix}
						\label{eqn:basic null space verification}
					\end{equation}
					where we have set $\tpose{\vec{u}_{\ast}} = \BB{0, \tpose{\vec{u}}}$.
					Clearly, $\vec{u}_{\ast} \in \setR^{m-1}$ and \eqref{eqn:basic null space verification} implies $\mat{X} \in \mathcal{N}_{0}\bb{m,n}$ by definition.
					By assumption, $\mat{X} \not \in \mathcal{M}\bb{m,n}$ and therefore $\mat{X} \in \mathcal{N}_{0}\bb{m,n} \setminus \mathcal{M}\bb{m,n}$.
				\end{enumerate}
	
				\textit{Induction Basis:}	We need to show that $\mathcal{N}\bb{\mathscr{S}_{\bb{3,n}}, 2} \setminus \mathcal{M}\bb{3,n} \subseteq \bb[\big]{\mathcal{N}_{0}\bb{3,n} \bigcup \mathcal{N}_{2}\bb{3,n}} \setminus \mathcal{M}\bb{3,n}$.
				Let $\mat{Z} \in \mathcal{N}\bb{\mathscr{S}_{\bb{3,n}}, 2} \setminus \mathcal{M}\bb{3,n}$.
				If $\mat{Z}\bb{1,:} \neq \tpose{\vec{0}}$ then it follows from \partname~\ref{itm:not zero} of the proof that $\mat{Z} \in \bb[\big]{\mathcal{N}_{0}\bb{3,n} \bigcup \mathcal{N}_{2}\bb{3,n}} \setminus \mathcal{M}\bb{3,n}$.
				Instead, if $\mat{Z}\bb{1,:} = \tpose{\vec{0}}$ then we have
				\begin{equation}
					\begin{bmatrix}
						0	\\
						\mathscr{S}_{\bb{2,n}}\bb[\big]{\mat{Z}\bb{2:3,:}}
					\end{bmatrix}
					=	\mathscr{S}_{\bb{2,n}}\bb{\mat{Z}}
					=	\vec{0}
				\end{equation}
				where the first equality is due to $\mathscr{S}_{\bb{2,n}}\bb{\cdot}$ summing elements along the anti-diagonals and the second equality follows from $\mat{Z} \in \mathcal{N}\bb{\mathscr{S}_{\bb{3,n}}, 2}$.
				Therefore, $\mat{Z}\bb{2:3,:} \in \mathcal{N}\bb{\mathscr{S}_{\bb{2,n}}, 2}$ implying that $\mat{Z}\bb{2:3,:} \in \mathcal{N}_{0}\bb{2,n}$ from \partname~\ref{sec:full null space proof part 2} of the proof.
				Thus, there exist $u \in \setR$ and $\vec{v} \in \setR^{n-1}$ such that
				\begin{equation}
					\mat{Z}\bb{2:3,:} =	\begin{bmatrix}
											u	&	0	\\
											0	&	-u
										\end{bmatrix}
										\begin{bmatrix}
											0	&	\tpose{\vec{v}}	\\
											\tpose{\vec{v}}	&	0
										\end{bmatrix}
				\end{equation}
				implying that
				\makeatletter
					\if@twocolumn
						\begin{equation}
							\begin{split}
								\mat{Z}	& =	\begin{bmatrix}
												\tpose{\vec{0}}	\\
												\mat{Z}\bb{2:3,:}
											\end{bmatrix}
										=	\begin{bmatrix}
												0	&	0	\\
												u	&	0	\\
												0	&	-u
											\end{bmatrix}
											\begin{bmatrix}
												0	&	\tpose{\vec{v}}	\\
												\tpose{\vec{v}}	&	0
											\end{bmatrix}	\\
										& =	\begin{bmatrix}
												\vec{u}_{\ast}	&	0	\\
												0	&	-\vec{u}_{\ast}
											\end{bmatrix}
											\begin{bmatrix}
												0	&	\tpose{\vec{v}}	\\
												\tpose{\vec{v}}	&	0
											\end{bmatrix}
							\end{split}
							\label{eqn:three dim null space verification}
						\end{equation}
					\else
						\begin{equation}
							\mat{Z}	=	\begin{bmatrix}
											\tpose{\vec{0}}	\\
											\mat{Z}\bb{2:3,:}
										\end{bmatrix}
									=	\begin{bmatrix}
											0	&	0	\\
											u	&	0	\\
											0	&	-u
										\end{bmatrix}
										\begin{bmatrix}
											0	&	\tpose{\vec{v}}	\\
											\tpose{\vec{v}}	&	0
										\end{bmatrix}
									=	\begin{bmatrix}
											\vec{u}_{\ast}	&	0	\\
											0	&	-\vec{u}_{\ast}
										\end{bmatrix}
										\begin{bmatrix}
											0	&	\tpose{\vec{v}}	\\
											\tpose{\vec{v}}	&	0
										\end{bmatrix}
							\label{eqn:three dim null space verification}
						\end{equation}
					\fi
				\makeatother
				where $\tpose{\vec{u}_{\ast}} = \BB{0,u}$.
				Thus, \eqref{eqn:three dim null space verification} implies that $\mat{Z} \in \mathcal{N}_{0}\bb{3,n}$.
				Since $\mat{Z} \not \in \mathcal{M}\bb{3,n}$ we have $\mat{Z} \in \mathcal{N}_{0}\bb{3,n} \setminus \mathcal{M}\bb{3,n}$.
	
			We have proved that if $\mat{X}\bb{m,1} \neq 0$ and $\mat{X} \in \mathcal{N}\bb{\mathscr{S}_{\bb{m,n}}, 2}$ then $\mat{X} \in \mathcal{N}_{0}\bb{m, n} \bigcup \mathcal{N}_{2}\bb{m, n}$.
			To finish the proof we need to consider the other scenario for $\mat{X} \not \in \mathcal{M}\bb{m,n}$, \ie~show that $\mat{X}\bb{1,n} \neq 0$ and $\mat{X} \in \mathcal{N}\bb{\mathscr{S}_{\bb{m,n}}, 2}$ imply $\mat{X} \in \mathcal{N}_{0}\bb{m, n} \bigcup \mathcal{N}_{2}\bb{m, n}$.
			The arguments for this case are ideologically similar to our approach for $\mat{X}\bb{m,1} \neq 0$ and hence we omit it.

	\section{Proof of \propositionname~\ref{prop:rank-2 nullspace exception set}}
		\label{sec:nullspace exception set proof}
		We explicitly define an $n-1$ dimensional parameterized set of matrices $\mathcal{M}_{2}\bb{n} \subset \setR^{n \times n}$ as follows,
		\makeatletter
			\if@twocolumn
				\begin{equation}
					\begin{split}
						\mathcal{M}_{2}\bb{n} & \triangleq	\mleft \{	\begin{bmatrix}
																			0	&	-\tpose{\vec{u}}	&	0	\\
																			\vec{u}	&	\mat{0}	&	\lambda \vec{u}	\\
																			0	&	-\lambda \tpose{\vec{u}}	&	0
																		\end{bmatrix} \, \middle|	\mright.	\\
						& \qquad \quad \mleft. \vphantom{\begin{bmatrix} 0 & -\tpose{\vec{u}} & 0 \\ \vec{u} & \mat{0} & \lambda \vec{u} \\ 0 & -\lambda \tpose{\vec{u}} & 0 \end{bmatrix}}
						\vec{u} \in \setR^{n-2} \setminus \cc{\vec{0}}, \lambda \in \setR \setminus \cc{0} \mright\}
					\end{split}
					\label{eqn:exception set}
				\end{equation}
			\else
				\begin{equation}
					\mathcal{M}_{2}\bb{n} \triangleq	\set{	\begin{bmatrix}
																	0	&	-\tpose{\vec{u}}	&	0	\\
																	\vec{u}	&	\mat{0}	&	\lambda \vec{u}	\\
																	0	&	-\lambda \tpose{\vec{u}}	&	0
																\end{bmatrix}}
															{\vec{u} \in \setR^{n-2} \setminus \cc{\vec{0}}, \lambda \in \setR \setminus \cc{0}}
					\label{eqn:exception set}
				\end{equation}
			\fi
		\makeatother
		and show that $\mathcal{M}_{2}\bb{n} \subseteq \mathcal{N}\bb{\mathscr{S}_{\bb{n,n}}, 2} \bigcap \mathcal{M}\bb{n,n} \setminus \bb[\big]{\mathcal{N}_{0}\bb{n, n} \bigcup \mathcal{N}_{2}\bb{n, n}}$.
		Let $\mat{X}$ be an arbitrary element of $\mathcal{M}_{2}\bb{n}$ and let $\bb{\vec{u} \in \setR^{n-2} \setminus \cc{\vec{0}}, \lambda \in \setR \setminus \cc{0}}$ denote some representative parameters certifying that $\mat{X} \in \mathcal{M}_{2}\bb{n}$ according to \eqref{eqn:exception set}.
		It can be easily verified that $\mat{X}$ is skew-symmetric.
		Since $\mathscr{S}_{\bb{n,n}}\bb{\cdot}$ sums elements along anti-diagonals (see \figurename~\ref{fig:anti-diagonal sum} for illustration), skew-symmetry of $\mat{X}$ implies that $\mat{X}$ is in the null space of $\mathscr{S}_{\bb{n,n}}\bb{\cdot}$.
		Further, $\mat{X}$ admits the factorization
		\begin{equation}
			\mat{X}	=	\begin{bmatrix}
							0	&	-\tpose{\vec{u}}	&	0	\\
							\vec{u}	&	\mat{0}	&	\lambda \vec{u}	\\
							0	&	-\lambda \tpose{\vec{u}}	&	0
						\end{bmatrix}
					=	\begin{bmatrix}
							0	&	1	\\
							\vec{u}	&	\vec{0}	\\
							0	&	\lambda
						\end{bmatrix}
						\begin{bmatrix}
							1	&	\tpose{\vec{0}}		&	\lambda	\\
							0	&	-\tpose{\vec{u}}	&	0
						\end{bmatrix}
			\label{eqn:exceptional representation}
		\end{equation}
		implying that $\rank{\mat{X}} \leq 2$ and therefore $\mat{X} \in \mathcal{N}\bb{\mathscr{S}_{\bb{n,n}}, 2}$.
		Since $\mat{X}(n,1) = \mat{X}(1,n) = 0$ we can also conclude that $\mat{X} \in \mathcal{N}\bb{\mathscr{S}_{\bb{n,n}}, 2} \bigcap \mathcal{M}\bb{n,n}$.

		To finish the proof, we need to show that $\mat{X} \not \in \mathcal{N}_{0}\bb{n, n} \bigcup \mathcal{N}_{2}\bb{n, n}$.
		We proceed for a proof by contradiction.
		First, assume that $\mat{X} \in \mathcal{N}_{2}\bb{n, n}$.
		Then there exist vectors $\vec{u}_{1}, \vec{u}_{2}, \vec{v}_{1}, \vec{v}_{2} \in \setR^{n-1}$ such that
		\begin{equation}
			\mat{X} =	\begin{bmatrix}
							\vec{u}_{1}	&	0	\\
							0	&	\vec{u}_{2}
						\end{bmatrix}
						\begin{bmatrix}
							0	&	\tpose{\vec{v}_{1}}	\\
							\tpose{\vec{v}_{2}}	&	0
						\end{bmatrix}.
			\label{eqn:regular representation}
		\end{equation}
		Since $\vec{u} \neq \vec{0}$ and $\lambda \neq 0$, \eqref{eqn:exceptional representation} implies $\mat{X}(1,:) \neq \tpose{\vec{0}}$ and $\mat{X}(n,:) \neq \tpose{\vec{0}}$.
		Now \eqref{eqn:regular representation} implies that $\mat{X}(1,:) = \BB{0, \vec{u}_{1}(1) \tpose{\vec{v}_{1}}}$ and $\mat{X}(n,:) = \BB{\vec{u}_{2}(n-1) \tpose{\vec{v}_{2}}, 0}$.
		Therefore, $\vec{u}_{1}(1) \neq 0$ and $\vec{u}_{2}(n-1) \neq 0$.
		Using the two different representations for $\mathcal{C}\bb{\mat{X}}$ in \eqref{eqn:exceptional representation} and \eqref{eqn:regular representation} we must have
		\begin{equation}
			\BB{0, \tpose{\vec{u}}, 0} = \gamma \BB{\tpose{\vec{u}_{1}}, 0} + \beta \BB{0, \tpose{\vec{u}_{2}}}
			\label{eqn:lin comb representations}
		\end{equation}
		for some $\tpose{\BB{\gamma, \beta}} \in \setR^{2} \setminus \cc{\vec{0}}$.
		However, \eqref{eqn:lin comb representations} implies $0 = \gamma \vec{u}_{1}(1)$ and $0 = \beta \vec{u}_{2}(n-1)$ and thus, $\gamma = \beta = 0$ since $\vec{u}_{1}(1) \neq 0$ and $\vec{u}_{2}(n-1) \neq 0$.
		This leads to a contradiction and therefore $\mat{X} \not \in \mathcal{N}_{2}\bb{n, n}$.
		Next, we assume $\mat{X} \in \mathcal{N}_{0}\bb{n, n}$.
		All of our arguments from the preceding case of $\mat{X} \in \mathcal{N}_{2}\bb{n, n}$ are still valid if we set $\vec{u}_{2} = -\vec{u}_{1}$.
		Therefore, once again we reach a contradiction and $\mat{X} \not \in \mathcal{N}_{0}\bb{n, n}$.
		So, we have $\mat{X} \not \in \mathcal{N}_{0}\bb{n, n} \bigcup \mathcal{N}_{2}\bb{n, n}$ and the proof is complete.

	\section{Proof of \lemmaname~\ref{lem:finite quotient set}}
		\label{sec:finite quotient set proof}
		If $\mathcal{Q}_{\sim}\bb{\vec{w}, d} = \emptyset$ then its cardinality is trivially upper bounded by $\bb{2d-2} \geq 2$.
		So, for the purpose of proving the cardinality upper bound, we assume that $\mathcal{Q}_{\sim}\bb{\vec{w}, d}$ is non-empty.
		Let $\bb{\vec{w}'_{\ast}, \gamma'} \in \mathcal{Q}_{\sim}\bb{\vec{w}, d}$ be arbitrarily selected.
		Since $0 \not \in \cc{\vec{w}(1), \vec{w}(d)}$ by assumption, we have $0 \not \in \cc{\vec{w}'_{\ast}(1), \vec{w}'_{\ast}(d-1), \sin \gamma', \cos \gamma'}$ by definition of $\mathcal{Q}_{\sim}\bb{\vec{w}, d}$, and therefore division by any element of the set $\cc{\vec{w}(1), \vec{w}(d), \vec{w}'_{\ast}(1), \vec{w}'_{\ast}(d-1), \sin \gamma', \cos \gamma'}$ is well defined.
		We provide a construction for $\bb{\vec{w}'_{\ast}, \gamma'}$ as follows.
		We have,
		\begin{equation}
			\cos \gamma' = \vec{w}(1)/\vec{w}'_{\ast}(1),	\quad
			\sin \gamma' = -\vec{w}(d)/\vec{w}'_{\ast}(d-1),
			\label{eqn:angle constraints}
		\end{equation}
		and for every $2 \leq j \leq d-1$,
		\begin{equation}
			\vec{w}(j) = \vec{w}'_{\ast}(j) \cos \gamma' - \vec{w}'_{\ast}(j-1) \sin \gamma'.
			\label{eqn:inverse mapping constraint}
		\end{equation}
		We define the vectors $\vec{w}/\vec{w}(d) = \vec{c} \in \setR^{d}$ and $\vec{w}'_{\ast}/\vec{w}'_{\ast}(d-1) = \vec{s} \in \setR^{d-1}$.
		For $j = 2,3,\dots,d-1$, we have
		\makeatletter
			\if@twocolumn
				\begin{subequations}
					\label{eqn:recursive defn}
					\begin{align}
						\vec{s}(j-1)
						& =	\frac{\vec{w}'_{\ast}(j-1)}{\vec{w}'_{\ast}(d-1)}	\notag \\
						& =	\frac{1}{\vec{w}'_{\ast}(d-1)}	\BB{\vec{w}'_{\ast}(j) \cot \gamma' - \vec{w}(j) \csc \gamma'}	\label{eqn:by substitution 1} \\
						& =	\frac{1}{\vec{w}'_{\ast}(d-1)} \Bd{-\vec{w}'_{\ast}(j) \frac{\vec{w}(1)}{\vec{w}(d)} \frac{\vec{w}'_{\ast}(d-1)}{\vec{w}'_{\ast}(1)}}	\notag \\
						& \hphantom{\frac{1}{\vec{w}'_{\ast}(d-1)}} \qquad
						+ \dB{\vec{w}(j) \frac{\vec{w}'_{\ast}(d-1)}{\vec{w}(d)}}	\label{eqn:by substitution 2} \\
						& =	\frac{\vec{w}(j)}{\vec{w}(d)} - \frac{\vec{w}(1)}{\vec{w}(d)} \frac{\vec{w}'_{\ast}(j)}{\vec{w}'_{\ast}(d-1)} \frac{\vec{w}'_{\ast}(d-1)}{\vec{w}'_{\ast}(1)}	\notag	\\
						& =	\vec{c}(j) -\vec{c}(1) \frac{\vec{s}(j)}{\vec{s}(1)}	\label{eqn:by definition}
					\end{align}
				\end{subequations}
			\else
				\begin{subequations}
					\label{eqn:recursive defn}
					\begin{align}
						\vec{s}(j-1)
						& =	\frac{\vec{w}'_{\ast}(j-1)}{\vec{w}'_{\ast}(d-1)}	\notag \\
						& =	\frac{1}{\vec{w}'_{\ast}(d-1)}	\BB{\vec{w}'_{\ast}(j) \cot \gamma' - \vec{w}(j) \csc \gamma'}	\label{eqn:by substitution 1} \\
						& =	\frac{1}{\vec{w}'_{\ast}(d-1)}	\BB{-\vec{w}'_{\ast}(j) \frac{\vec{w}(1)}{\vec{w}(d)} \frac{\vec{w}'_{\ast}(d-1)}{\vec{w}'_{\ast}(1)} + \vec{w}(j) \frac{\vec{w}'_{\ast}(d-1)}{\vec{w}(d)}}	\label{eqn:by substitution 2} \\
						& =	\frac{\vec{w}(j)}{\vec{w}(d)} - \frac{\vec{w}(1)}{\vec{w}(d)} \frac{\vec{w}'_{\ast}(j)}{\vec{w}'_{\ast}(d-1)} \frac{\vec{w}'_{\ast}(d-1)}{\vec{w}'_{\ast}(1)}	\notag	\\
						& =	\vec{c}(j) -\vec{c}(1) \frac{\vec{s}(j)}{\vec{s}(1)}	\label{eqn:by definition}
					\end{align}
				\end{subequations}
			\fi
		\makeatother
		where \eqref{eqn:by substitution 1} follows from \eqref{eqn:inverse mapping constraint}, \eqref{eqn:by substitution 2} follows from \eqref{eqn:angle constraints} using basic trigonometric relationships, and \eqref{eqn:by definition} follows from the definitions for vectors $\vec{c}$ and $\vec{s}$.
		Equation \eqref{eqn:recursive defn} represents a set of constraints on the vector variable $\vec{s}$ with $\vec{c}$ known.
		Since $\vec{s}(d-1) = 1$ by definition, \eqref{eqn:recursive defn} can be solved recursively for $\vec{s}(d-2), \vec{s}(d-3), \dots, \vec{s}(1)$ (in that order) as polynomial expressions in the variable $1/\vec{s}(1)$.
		Specifically, for $j=2$, \eqref{eqn:recursive defn} leads to an expression of $\vec{s}(1)$ as a $(d-2)$ degree polynomial of the variable $1/\vec{s}(1)$ and therefore, represents a consistency equation that must be satisfied by every solution to \eqref{eqn:recursive defn} (\ie~by every element of the set $\mathcal{Q}_{\sim}\bb{\vec{w}, d}$).
		Furthermore, dividing the consistency equation by $\vec{s}(1)$ on both \lhs~and \rhs~results in a polynomial equation of degree $(d-1)$ in the variable $1/\vec{s}(1)$ and therefore $\vec{s}(1)$ can admit at most $(d-1)$ distinct values.
		If $\mathcal{S}$ denotes the set of admissible values of $\vec{s}(1)$ then $\card{\mathcal{S}} \leq \bb{d-1}$.
		For a given value of $\vec{s}(1) \in \mathcal{S}$, \eqref{eqn:recursive defn} uniquely determines $\vec{s}$ since it gives a recepie to compute $\vec{s}(j)$ for every $2 \leq j \leq d-2$ and we have $\vec{s}(d-1) = 1$ by definition.
		From \eqref{eqn:angle constraints}, we get
		\begin{equation}
			\tan \gamma' = -\frac{\vec{w}(d)}{\vec{w}(1)} \frac{\vec{w}'_{\ast}(1)}{\vec{w}'_{\ast}(d-1)} = -\frac{\vec{s}(1)}{\vec{c}(1)}
			\label{eqn:tan finite cardinality}
		\end{equation}
		implying that given a value of $\vec{s}(1) \in \mathcal{S}$, $\tan \gamma'$ is uniquely determined, and $\csc \gamma' \in \cc{\pm \sqrt{1 + \tan^{-2} \gamma'}}$ and $\gamma' \in \setA$ each admits at most two distinct values.
		Therefore, the pair of scalars $\bb{\vec{s}(1), \gamma'}$ admits at most $2 \card{\mathcal{S}}$ distinct values.
		Using \eqref{eqn:angle constraints} and the definition of $\vec{s}$, we have $\vec{w}'_{\ast} = \vec{w}'_{\ast}(d-1) \cdot \vec{s} = -\vec{s} \cdot \vec{w}(d) \cdot \csc \gamma'$ implying that $\vec{w}'_{\ast}$ is uniquely determined by $\bb{\vec{s}(1), \gamma'}$.
		Therefore, $\bb{\vec{w}'_{\ast}, \gamma'}$ admits at most $2\card{\mathcal{S}} \leq \bb{2d-2}$ distinct values.
		Since $\bb{\vec{w}'_{\ast}, \gamma'}$ represents an arbitrary element of $\mathcal{Q}_{\sim}\bb{\vec{w}, d}$, we conclude that $\mathcal{Q}_{\sim}\bb{\vec{w}, d}$ is finite with cardinality at most $\bb{2d-2}$.
		As a boundary case, notice that when $d = 2$, \eqref{eqn:inverse mapping constraint} and \eqref{eqn:recursive defn} are vacuous and \eqref{eqn:angle constraints} suffices to prove the result since it implies the existence of exactly two elements in $\mathcal{Q}_{\sim}\bb{\vec{w}, 2}$, \viz~$\bb{\vec{w}'_{\ast}, \gamma'}$ and $\bb{-\vec{w}'_{\ast}, 2\pi - \gamma'}$.

		To prove that $\mathcal{Q}_{\sim}\bb{\vec{w}, d}$ is non-empty when $d$ is an even integer, we proceed as follows.
		Given $\vec{w} \in \setR^{d}$ with $0 \not \in \cc{\vec{w}(1), \vec{w}(d)}$, $\vec{w}/\vec{w}(d) = \vec{c} \in \setR^{d}$ is well defined, admits $\vec{c}(1) = \vec{w}(1)/\vec{w}(d) \neq 0$ and is completely known.
		Let $\vec{s} \in \setR^{d-1}$ be a vector variable such that it satisfies \eqref{eqn:by definition} for $j = 2,3,\dots,d-1$ and admits $\vec{s}(d-1) = 1$.
		If such a vector $\vec{s}$ could be found with $\vec{s}(1) \neq 0$, then using \eqref{eqn:tan finite cardinality} to choose $\gamma' \in \setA$ and setting $\vec{w}'_{\ast} = -\vec{s} \cdot \vec{w}(d) \cdot \csc \gamma'$ yields and element $\bb{\vec{w}'_{\ast}, \gamma'} \in \mathcal{Q}_{\sim}\bb{\vec{w}, d}$, since
		\begin{enumerate}
			\item	\eqref{eqn:inverse mapping constraint} follows from being able to reverse the sequence of implications in the preceding part of the proof, \ie~$\eqref{eqn:by definition} \implies \eqref{eqn:by substitution 2} \implies \eqref{eqn:by substitution 1} \implies \eqref{eqn:inverse mapping constraint}$ for every $2 \leq j \leq d-1$,
			\item	\eqref{eqn:angle constraints} follows directly from \eqref{eqn:tan finite cardinality}, $\vec{w}'_{\ast} = -\vec{s} \cdot \vec{w}(d) \cdot \csc \gamma'$ and $\vec{s}(d-1) = 1$, and
			\item	$0 \not \in \cc{\vec{w}(1), \vec{w}(d)}$ is satisfied since $0 \not \in \cc{\vec{s}(1), \vec{c}(1)}$ and $\tan \gamma' \not \in \cc{0, \pm \infty}$.
		\end{enumerate}
		For the boundary case of $d = 2$, \eqref{eqn:inverse mapping constraint} is vacuous and $\vec{w}'_{\ast}(1) = \sqrt{\vec{w}^{2}(1) + \vec{w}^{2}(2)}$ satisfies \eqref{eqn:angle constraints} for $\gamma' = \arccos\bb{\vec{w}(1)/\vec{w}'_{\ast}(1)}$, thus yielding an element in $\mathcal{Q}_{\sim}\bb{\vec{w}, 2}$ and proving the result.
		For $d \geq 3$, we follow the recursive solution process for \eqref{eqn:by definition} to arrive at a consistency equation for $\vec{s}(1)$ as previously described, \ie~we recursively express $\vec{s}(d-2), \vec{s}(d-3), \dots, \vec{s}(1)$ (in that order) as polynomials in the variable $1/\vec{s}(1)$ leading to a polynomial expression for $\vec{s}(1)$ of degree $(d-2)$ in the variable $1/\vec{s}(1)$.
		This further leads to a polynomial equation of degree $(d-1)$ in the variable $1/\vec{s}(1)$ with a non-zero constant term and hence admits at least one non-zero real root whenever $\bb{d-1}$ is an odd integer (equivalently, if $d$ is even).
		Using this value of $1/\vec{s}(1)$, \eqref{eqn:by definition} uniquely determines the vector $\vec{s}$ since it provides a recepie to recursively compute $\vec{s}(j)$ for every $2 \leq j \leq d-2$ and we have $\vec{s}(d-1) = 1$ by definition.
		This demonstrates the existence of the vector $\vec{s} \in \setR^{d-1}$ with the desired characteristics and completes the proof.

	\section{Proof of \theoremname~\ref{thm:ae unident}}
		\label{sec:ae unidentifiability proof}
		As in the statement of the theorem, let $\bb{\vec{x}, \vec{y}} \in \mathcal{K}$ be an arbitrary vector pair with $\mathcal{K}$ defined by \eqref{eqn:non-sparse domain}.
		Since $m,n \geq 4$ are even integers, invoking \lemmaname~\ref{lem:finite quotient set} once each for $\vec{x}$ and $\vec{y}$, we can construct vectors $\vec{u} \in \setR^{m-1}$, $\vec{v} \in \setR^{n-1}$ and scalars $\theta, \phi \in \setR$ such that
		\begin{equation}
			\vec{x} =	\begin{bmatrix}
							\vec{u}	&	0	\\
							0	&	-\vec{u}
						\end{bmatrix}
						\begin{bmatrix}
							\cos \theta	\\
							\sin \theta
						\end{bmatrix},
			\quad
			\vec{y} =	\begin{bmatrix}
							0	&	\vec{v}	\\
							\vec{v}	&	0
						\end{bmatrix}
						\begin{bmatrix}
							\sin \phi	\\
							-\cos \phi
						\end{bmatrix}.
			\label{eqn:decomposition equation}
		\end{equation}
		Equation \eqref{eqn:decomposition equation} implies that the dependence of $\vec{y}$ is uniformly continuous on the $n$ real numbers $\bb{\phi, \vec{v}(1), \vec{v}(2), \dots, \vec{v}(n-1)}$ (with a non-singular Jacobian matrix) and these $n$ real numbers completely parametrize $\vec{y}$.
		Further, since the measure over $\vec{y}$ is absolutely continuous \wrt~the $n$ dimensional Lebesgue measure, it is possible to choose a measure over $\phi$ that is absolutely continuous \wrt~the one dimensional Lebesgue measure, and therefore $\phi \not \in \set{\bb{l+1/2}\pi}{l \in \setZ}$ is true almost everywhere \wrt~the measure over $\phi$.
		Finally, since $\vec{x}$ and $\vec{y}$ have disjoint parameterizations, $\phi - \theta \not \in \set{l\pi}{l \in \setZ}$ is true almost everywhere \wrt~the measure over $\phi$ (owing to absolute continuity \wrt~one dimensional Lebesgue measure).
		Thus, $\phi \not \in \set{\bb{l+1/2}\pi, \theta + l\pi}{l \in \setZ}$ is true almost everywhere \wrt~the measure over $\phi$ and therefore also \wrt~the measure over $\vec{y}$.

		Once we have vectors $\vec{u}$ and $\vec{v}$, we consider the decomposition
		\begin{equation}
			\begin{split}
				\mat{X}	& =	\begin{bmatrix}
								\vec{u}	&	0	\\
								0	&	-\vec{u}
							\end{bmatrix}
							\begin{bmatrix}
								\cos \theta	\\
								\sin \theta
							\end{bmatrix}
							\begin{bmatrix}
								\sin \phi	&	-\cos \phi
							\end{bmatrix}
							\begin{bmatrix}
								0		&	\tpose{\vec{v}}	\\
								\tpose{\vec{v}}	&	0
							\end{bmatrix}	\\
						& {} +	\begin{bmatrix}
									\vec{u}	&	0	\\
									0	&	-\vec{u}
								\end{bmatrix}
								\begin{bmatrix}
									\cos \phi	\\
									\sin \phi
								\end{bmatrix}
								\begin{bmatrix}
									-\sin \theta	&	\cos \theta
								\end{bmatrix}
								\begin{bmatrix}
									0		&	\tpose{\vec{v}}	\\
									\tpose{\vec{v}}	&	0
								\end{bmatrix}	\\
						& =	\sin\bb{\phi - \theta}	\begin{bmatrix}
														\vec{u}	&		0	\\
														0	&	-\vec{u}
													\end{bmatrix}
													\begin{bmatrix}
														0	&	\tpose{\vec{v}}	\\
														\tpose{\vec{v}}	&	0
													\end{bmatrix}
			\end{split}
			\label{eqn:adversarial decomposition}
		\end{equation}
		for $\mat{X} \in \mathcal{N}\bb{\mathscr{S}, 2}$, as in \eqref{eqn:rank-2 nullspace}.
		Setting
		\begin{alignat}{2}
			\vec{x}'	& =	\begin{bmatrix}
								\vec{u}	&	0	\\
								0	&	-\vec{u}
							\end{bmatrix}
							\begin{bmatrix}
								\cos \phi	\\
								\sin \phi
							\end{bmatrix}, %
						& \quad %
			\vec{y}'	& =	\begin{bmatrix}
								0	&	\vec{v}	\\
								\vec{v}	&	0
							\end{bmatrix}
							\begin{bmatrix}
								\sin \theta	\\
								-\cos \theta
							\end{bmatrix}
			\label{eqn:adversarial pair}
		\end{alignat}
		and observing \eqref{eqn:decomposition equation} and \eqref{eqn:adversarial decomposition} we conclude that $\vec{x}\tpose{\vec{y}} - \vec{x}'\tpose{\bb{\vec{y}'}} = \mat{X} \in \mathcal{N}\bb{\mathscr{S}, 2}$ and therefore the pairs $\bb{\vec{x}, \vec{y}}$ and $\bb{\vec{x}', \vec{y}'}$ produce the same convolved output.
		Since $\vec{x}(1) \neq 0$, \eqref{eqn:decomposition equation} implies that $\vec{u}(1) \neq 0$ and therefore \eqref{eqn:adversarial pair} implies $\vec{x}'(1) \neq 0$ for $\phi \not \in \set{\bb{l+1/2}\pi}{l \in \setZ}$.
		So, assuming $\phi \not \in \set{\bb{l+1/2}\pi}{l \in \setZ}$ implies that $\bb{\vec{x}', \vec{y}'} \in \mathcal{K}$.
		Further, $\vec{x}$ and $\vec{x}'$ are linearly independent if $\phi - \theta \not \in \set{l\pi}{l \in \setZ}$, implying that $\bb{\vec{x}, \vec{y}}$ is unidentifiable by \definitionname~\ref{defn:identifiability} with $\bb{\vec{x}', \vec{y}'}$ as the certificate.
		The proof is complete since $\phi \not \in \set{\bb{l+1/2}\pi, \theta + l\pi}{l \in \setZ}$ is true almost everywhere \wrt~the measure over $\vec{y}$.

\end{document}